\documentclass[journal]{IEEEtran}
\usepackage{multirow}

\usepackage{cite}
\usepackage{url}

\ifCLASSINFOpdf
  \usepackage[pdftex]{graphicx}
  \DeclareGraphicsExtensions{.pdf,.jpeg,.png}
\else
\fi
\usepackage{array}
\ifCLASSOPTIONcompsoc
  \usepackage[caption=false,font=normalsize,labelfont=sf,textfont=sf]{subfig}
\else
  \usepackage[caption=false,font=footnotesize]{subfig}
\fi

\usepackage[english]{babel}
\usepackage{amsmath,amssymb,bbm}
\usepackage{algorithm}
\usepackage[noend]{algpseudocode}
\usepackage{lipsum}
\usepackage{color}

\newcommand{\assign}{:=}
\newcommand{\nobracket}{}

\newcommand{\nosymbol}{}
\newcommand{\tmmathbf}[1]{\ensuremath{\boldsymbol{#1}}}
\newcommand{\tmop}[1]{\ensuremath{\operatorname{#1}}}

\newcolumntype{L}[1]{>{\raggedright\let\newline\\\arraybackslash\hspace{0pt}}m{#1}}
\newcolumntype{C}[1]{>{\centering\let\newline\\\arraybackslash\hspace{0pt}}m{#1}}
\newcolumntype{R}[1]{>{\raggedleft\let\newline\\\arraybackslash\hspace{0pt}}m{#1}}

\title{Fast Discrete Distribution Clustering Using Wasserstein Barycenter with Sparse Support}
\author{Jianbo~Ye, Panruo~Wu,
        James~Z.~Wang
        and~Jia~Li~\IEEEmembership{Senior Member,~IEEE}% <-this % stops a space
\thanks{J. Ye and J. Z. Wang are with College of Information Sciences and Technology,
The Pennsylvania State University, University Park, PA 16802, USA  
(emails: \textrm{\{jxy198, jwang\}@ist.psu.edu}).}% <-this % stops a space
\thanks{J. Li is with Department of Statistics, 
The Pennsylvania State University, University Park, PA, 16802, USA
(email: \textrm{jiali@stat.psu.edu}).}
\thanks{P. Wu is with Department of Computer Science and Engineering, 
University of California, Riverside, CA 92521, USA
(email: \textrm{pwu011@cs.ucr.edu}).}% <-this % stops a space
}

\begin{document}
% for an overview, comment it out for submission
%\tableofcontents

\maketitle
\setcounter{page}{1}

\begin{abstract}
In a variety of research areas, the weighted bag of vectors and the histogram are
widely used descriptors for complex objects. Both can be expressed as
discrete distributions.  D2-clustering pursues the minimum total 
within-cluster variation for a set of discrete distributions subject to the
Kantorovich-Wasserstein metric. D2-clustering has a severe scalability issue,
the bottleneck being the computation of a centroid distribution, called Wasserstein barycenter, 
that minimizes its sum of squared distances to the cluster members. In this paper, 
we develop a modified Bregman ADMM approach
for computing the approximate discrete Wasserstein barycenter of large clusters. In the case when the support points of the barycenters are unknown and have low cardinality, our method achieves high accuracy empirically at a much reduced computational cost.
The strengths and weaknesses of our method and its alternatives are examined through experiments, and we recommend
scenarios for their respective usage.  
Moreover, we develop both serial and parallelized versions of
the algorithm. By experimenting with large-scale data, we demonstrate the computational efficiency
of the new methods and investigate their convergence properties and numerical
stability. The clustering results obtained on several datasets in different domains
are highly competitive in comparison with some widely used
methods in the corresponding areas.
\end{abstract}


% \fi
\IEEEpeerreviewmaketitle

\section{Introduction}
The discrete distribution, or discrete probability measure, is a well-adopted and succinct 
way to summarize a batch of data. It often serves as a descriptor for 
complex instances encountered in machine learning, {\it e.g.}, images,
sequences, and documents, where each instance itself is converted to a data collection instead of a vector.  
In a variety of research areas including multimedia
retrieval and document analysis, the celebrated bag of
``words'' data model is intrinsically a discrete distribution.  
The widely used normalized histogram is
a special case of discrete distributions with a fixed set of support points across the instances.  In many applications involving moderate or high dimensions, the number of bins in the histograms is enormous because of the diversity across instances, while the histogram for any individual instance is highly sparse.  This distribution pattern is frequently observed for
collections of images and text documents.  The discrete distribution can function
as a sparse representation for the histogram, where support points and their
probabilities are specified in pairs, eliminating the
restriction of a fixed quantization codebook across instances.

The goal of this paper is to develop computationally efficient algorithms
for clustering discrete distributions under the Wasserstein distance.  
The Wasserstein distance is a true metric for measures\cite{rachev1984} 
and can be traced back to the mass transport problem proposed by Monge
in the 1780s and the relaxed formulation by Kantorovich in the 1940s~\cite{villani2003topics}. 
Mallows\cite{mallows1972} used this distance to prove some
theoretical results in statistics, and thus the name Mallows 
distance has been used by statisticians.  
In computer science, this formulation is better known as the
Earth Mover's Distance \cite{rubner2000earth}. In signal processing, 
it is closely related to the Optimal Sub-Pattern Assignment (OSPA) 
distance~\cite{schuhmacher2008consistent} used recently for 
multi-target tracking~\cite{baum2015wasserstein}. 
A more complete account on the history of this distance can be found in \cite{villani2008optimal}.
The Wasserstein distance is computationally expensive
because it has no closed form solution and
{can only be solved with a worst case time complexity} $O(m^3\log m)$ subject to 
the number of support points in the distribution~\cite{orlin1993faster}.

We adopt the well-accepted criterion of minimizing the total within-cluster 
variation under the Wasserstein distance, similarly as Lloyd's 
K-means for vectors under the Euclidean distance.  This clustering problem was originally explored by Li and Wang \cite{li2008real}, who coined the phrase {\em D2-clustering}, 
referring to both the optimization problem and their particular algorithm. 
%We will follow their terminology, but the specific meaning should be clear from context. 
Motivations for using the Wasserstein
distance in practice are strongly argued by researchers ({\it e.g.} ~\cite{rubner2000earth,li2008real,pele2009fast,cuturi2013sinkhorn,cuturi2014fast}), 
and its theoretical significance is well supported in the optimal transport (OT) literature 
(see the seminal books of Villani \cite{villani2008optimal,villani2003topics}). 
D2-clustering computes a principled centroid to summarize each cluster of data.  
The centroids computed from those clusters---also represented by discrete
distributions---are highly valuable for subsequent learning or data mining tasks.
Although D2-clustering holds much promise because of the advantages of the Wasserstein distance
and its kinship with K-means, the high computational complexity 
has limited its applications in large-scale learning.  

\subsection{Our Contributions} 
We have developed an efficient and robust method for 
optimizing the centroids in D2-clustering based on a modified version of the Bregman ADMM (B-ADMM).  
The data setting in D2-clustering has subtle differences from those taken by the previous OT literature on computing barycenters.
As a result, the modified B-ADMM approach is currently the 
most suitable (yet approximate) method for D2-clustering, surpassing
the other Wasserstein barycenter approaches adopted into the clustering process, namely: 
subgradient based method, ADMM, and entropy regularization. 
The properties of the modified B-ADMM approach are thoroughly examined via experiments,
and the new method called {\em AD2-clustering} is capable of handling a large number of instances,
provided the support size of each instance is limited to several hundred. 
We have also developed a parallel algorithm for the modified B-ADMM method in 
a multi-core environment with adequate scaling efficiency subject to hundreds of CPUs. 
Finally, we tested our new algorithm on several large-scale real-world datasets, 
revealing the high competitiveness of AD2-clustering. 

\subsection{Related Work}
\noindent\textbf{Clustering distributions:}
Distribution clustering can be subjected to different affinity definitions. For example,
Bregman clustering pursues the minimal distortion between the cluster prototype,
called the Bregman representative, and cluster members
according to a certain Bregman divergence~\cite{banerjee2005clustering}.
In comparison, D2-clustering is an extension of K-means to discrete distributions 
under the Wasserstein distance~\cite{li2008real}, and the cluster prototype
is an approximate Wasserstein barycenter with a sparse finite support set.  
In the D2-clustering framework, solving the cluster prototype or the centroid for discrete distributions under the 
Wasserstein distance is computationally challenging \cite{cuturi2014fast,ye2014scaling,zhang2015parallel}.
In order to scale up the computation of D2-clustering,
a divide-and-conquer approach has been proposed~\cite{zhang2015parallel}, 
but the method is ad-hoc from an optimization perspective. 
A standard ADMM approach has also been explored~\cite{ye2014scaling}, 
but its efficiency is still inadequate for large datasets.
Although fast computation of Wasserstein distances has been much explored \cite{pele2009fast,cuturi2013sinkhorn,wang2014bregman}, how to perform top-down 
clustering efficiently based on the distance has not. 
We present below the related work and discuss their relationships with our current work. \\

\noindent\textbf{The true discrete Wasserstein barycenter and two approximation strategies:}
The centroid of a collection of distributions minimizing the average $p$th-order power of the $L_p$ Wasserstein distance is called Wasserstein barycenter~\cite{agueh2011barycenters}. In the D2-clustering algorithm~\cite{li2008real}, the 2nd-order Wasserstein barycenter is simply referred to as a prototype or centroid, and is solved for the case of an unknown support with a pre-given cardinality.
The existence, uniqueness, regularity and other properties of the 2nd-order Wasserstein barycenter have been established mathematically for continuous measures in the Euclidean space~\cite{agueh2011barycenters}. The situation for discrete distributions, however, 
is more intricate, as will be explained later. 

Given $N$ arbitrary discrete distributions each with $\bar{m}$ support points, 
their true Wasserstein barycenter in theory can be solved via linear programming~\cite{agueh2011barycenters,anderes2015discrete}.
This approach is logical because the support points of the Wasserstein barycenter 
can only locate at a finite (yet huge) number of possible positions. Yet,
solving the true discrete barycenter quickly becomes intractable
even for a rather small number of distributions containing only 10 support points each.
Anderes {\it et al.}~\cite{anderes2015discrete} made important theoretical progress on this particular challenge by proving that 
the actual support of a true barycenter of $N$ such distributions is extremely sparse, with cardinality $m$ no greater than $\bar{m}N$. Unfortunately, the complexity of the problem is not reduced practically because so far there is no theoretically ensured way to sift out the optimal sparse locations. Anderes {\it et al.}'s approach, though, does backup
the practice of assuming a pre-selected number of support 
points in a barycenter as an approximation to the true solution.

To achieve good approximation, two computational strategies are useful in an optimization framework. 
\begin{enumerate}
\item[(i)] Carefully select beforehand a large and representative set of support points as an approximation to the support of the true barycenter ({\it e.g.}, K-means).
\item[(ii)] Allow the support points in a barycenter to adjust positions at 
every $\tau$ iterations.
\end{enumerate} 
The first strategy of fixing the support of a barycenter can yield 
adequate approximation quality in low dimensions ({\it e.g.} 1D/2D histogram data)~\cite{cuturi2014fast,benamou2014iterative}, but can face the challenge of an exponentially growing support size when the dimension increases.
The second strategy allows one to use a possibly much smaller 
number of support points in a barycenter to achieve the same 
level of accuracy~\cite{li2008real,zhang2015parallel,ye2014scaling,cuturi2014fast}.
Because the time complexity per iteration of existing iterative methods is $O(\bar{m}m N)$, a smaller $m$ can also save much computation,
and the extra amount of time $O(\bar{m} m d N / \tau)$ can be used to recalculate the distance matrices. 
In the extreme case when the barycenter support size is set to one ($m=1$), 
D2-clustering reduces to K-means on the distribution means, which is a meaningful way of data reduction in its own right. Our experiments indicate that in practice
a large $m$ in D2-clustering is usually unnecessary (see Section~\ref{sec:expr} for related discussions). 

In applications on high-dimensional data, optimizing the support points is preferred to fixing them from the beginning. This option, however, leads to a non-convex optimization problem. Our work aims at developing practical numerical methods. In particular, our method optimizes jointly the locations and weights of the support points in a single loop without resorting to a bi-level optimization reformulation, as was done in earlier work~\cite{li2008real,cuturi2014fast}.\\

\noindent\textbf{Solving discrete Wasserstein barycenter in different data settings:}
Recently, a series of works have been devoted to solving 
the Wasserstein barycenter given a set of distributions ({\it e.g.} ~\cite{carlier2014numerical,cuturi2014fast,benamou2014iterative,ye2014scaling,cuturi2015smoothed}). How our method compares with the existing ones depends strongly on the specific data setting. We discuss the comparisons in details below and
promote the use of our new method, AD2-clustering.

%Several algorithms including ours involve iterative updates of variables 
%in a probability simplex, yielding similar numerical iterations \cite{benamou2014iterative,wang2014bregman,cuturi2014fast} with the same time complexity. 
%They can be all viewed as variants of the iterative proportional fitting procedure (IPFP)
%tracing back to the early 1940s~\cite{deming1940least,bacharach1965estimating}. 
In \cite{benamou2014iterative,cuturi2014fast,cuturi2013sinkhorn}, novel algorithms have been developed for solving the Wasserstein barycenter by adding an entropy regularization term on the optimal transport matching weights. The regularization is imposed on the transport weights, but not the barycenter distribution.
%Using the entropic schemes for OT has a long tradition in linear 
%programming literature~\cite{carlier2015convergence}.
In particular, iterative Bregman projection (IBP)~\cite{benamou2014iterative} 
can solve an approximation to the Wasserstein barycenter. 
IBP is highly memory efficient for distributions with a shared support set ({\it e.g.} histogram data), with a memory complexity $O((m + \bar{m})N)$.
In comparison, our modified B-ADMM approach is of the same time complexity, but requires $O(m\bar{m}N)$ memory. If $N$ is large, memory constraints can limit our approach
to problems with relatively small $m$ or $\bar{m}$. While the first strategy may not meet the memory constraint, 
the second approximation strategy is crucial for reaching high accuracy with our approach.  
Conventional OT literature emphasizes computing the Wasserstein barycenter for a small number of 
 instances with dense representations 
({\it e.g.}~\cite{solomon2015convolutional,rabin2011wasserstein}); and IBP is more suitable.
Yet in many machine learning and signal processing applications, each instance is 
represented by a discrete distribution with a sparse finite support set ({\it i.e.}, $\bar{m}$ is small). 
The memory limitation of B-ADMM can be avoided via parallelization until the time allocation is spent. Our focus is thus to achieve scalability in $N$. 

As demonstrated by experiments, B-ADMM has advantages over IBP that motivate its usage in our algorithm. 
If the distributions do not share the support set, 
IBP has the same memory complexity 
$O(m\bar{m}N)$ (for caching the distance matrix per instance) as our approach has.
In addition, B-ADMM~\cite{wang2014bregman}, based on which our approach is developed, has the following advantages: (1) It yields the exact OT and distance in the limit of iterations. Note that the ADMM parameter 
does not trade off the convergence rate. (2) It requires little tuning of hyper-parameters and easily accommodates warm starts (to be illustrated later), which are valuable traits for the task of D2-clustering. (3) It works well with single-precision floats, and thus it is not restricted by the machine precision constraint. In contrast, IBP requires more tuning and may encounter precision overflow which is hard to address. Our experiments show that when
its coupling solutions are used, the resulting discrete
Wasserstein barycenters with sparse finite support sets are not as accurate as those by 
B-ADMM (see~\cite{benamou2014iterative} and our experiments).%
\footnote{Here ``accurate'' means close to a local 
minimizer of sum of the (squared) Wasserstein distances.}\\

\noindent\textbf{Optimization methods revisited:}
Our main algorithm is inspired by the B-ADMM algorithm of 
Wang and Banerjee~\cite{wang2014bregman} for solving OT fast. 
They developed the two-block variant of 
ADMM~\cite{boyd2011distributed} along with Bregman divergence to solve OT 
when the number of support points is extremely large.
Its algorithmic relation to IBP~\cite{benamou2014iterative} is discussed in Section~\ref{sec:compare}.
The OT problem at a moderate scale can in fact be efficiently handled 
by state-of-the-art LP solvers~\cite{tang2013earth}.
As demonstrated by the line of work on solving the barycenter, 
optimizing the Wasserstein barycenter is rather different from computing the distance. 
Whereas na\"ively adapting the B-ADMM to Wasserstein barycenter does not result in a proper algorithm,
our modified B-ADMM algorithm effectively addresses the Wasserstein barycenter problem.
The modification we made on B-ADMM is necessary.
Although the modified B-ADMM approach is not guaranteed to 
converge to a local optimum, it often
yields a solution very close to the local optimum. 
The new method is shown empirically to achieve higher accuracy than 
IBP or its derivatives when distributions are supported on finite sets. 

Finally, we note that although solving a single barycenter for a fixed set is 
a key component in D2-clustering, the task of clustering per se bears some extra technical subtleties. 
In a clustering setup, the partition of samples varies over the iterations, 
and a sequence of Wasserstein barycenters are solved.
We found that the robustness with respect to the hyper-parameters in the optimization 
algorithms is as important as the speed of solving one centroid because 
tuning these parameters over many iterations of different partitions is impractical. \\

\noindent\textbf{Outline:} The rest of the paper is organized as follows.
In Section~\ref{sec:d2_pre}, we define notations and provide preliminaries on 
D2-clustering. In Section~\ref{sec:centroid}, we develop two scalable optimization approaches for the
Wasserstein barycenter problem and explain how they are embedded in the overall algorithm for D2-clustering.  
In Section~\ref{sec:compare}, our main algorithm is compared with the IBP approach. 
In Section~\ref{sec:impl}, several implementation issues, including initialization, additional techniques for speed-up, and parallelization, are addressed. Comparisons of the algorithms in complexity/performance and guidelines for usage in practice are provided in Section~\ref{sec:compl}.
Section~\ref{sec:expr} reports experimental results. The numerical properties and computing performance of the algorithms are investigated; and clustering results on multiple datasets from different domains are compared with those obtained by widely used methods in the respective areas.  
Finally, we conclude in Section~\ref{sec:concl} and discuss the limitations of the current work.

\section{Discrete Distribution Clustering}\label{sec:d2_pre}
%\subsection{Preliminaries}\label{sec:prel}
Consider discrete distributions with sparse finite support specified by a set of
support points and their associated probabilities, a.k.a. weights: 
$$\{ ( w_{1} ,x_{1} ) , \ldots , ( w_{m} ,x_{m} ) \},$$
where $\sum_{i=1}^{m} w_{i} =1$ with $w_{i} \geqslant 0$, and $x_{i} \in
\mathbbm{M}$ for $i=1, \ldots ,m$. 
%Typically, we assume $m$ is no more than several hundreds.
%We note that this formulation includes the
%empirical estimation of the distribution where probability $w_{i} =
%\frac{1}{m}$ is assigned to each point in an i.i.d. sample of size $m$. 
Usually, $\mathbbm{M}=\mathbbm{R}^{d}$ is the
$d$-dimensional Euclidean space with the $L_p$ norm, and $x_i$'s are also called support vectors. 
$\mathbbm{M}$ can also be a symbolic set provided with
symbol-to-symbol dissimilarity. 
The Wasserstein distance between distributions $P^{( a )}=\{(w_i^{(a)},
x_i^{(a)}), i=1, ..., m_a\}$ and $P^{( b )}=\{(w_i^{(b)}, x_i^{(b)}),
i=1, ..., m_b\}$ is solved by the following linear programming (LP).  For notation brevity, let $c(x_i^{(a)}, x_j^{(b)})=\|x_i^{(a)}-x_j^{(b)}\|_p^p$. 
Define index set $\mathcal{I}_a=\{1, ..., m_a\}$ and  $\mathcal{I}_b$
likewise. We define $\left(W_p( P^{( a )} ,P^{( b )})\right)^{p}\assign$
\begin{equation}
\begin{array}{rl}
 \min\limits_{\{\pi_{i,j} \geqslant 0\}} &
  \sum\limits_{i\in\mathcal{I}_a,j\in\mathcal{I}_b} \pi_{i,j}  c ( x_{i}^{( a )} ,x_{j}^{( b )} )\; , \\
   \textrm{s.t.}\;\; & \sum_{i=1}^{m_a} \pi_{i,j}  =w_{j}^{( b )} ,\; \forall
   j\in\mathcal{I}_b\; ,\\ & \sum_{j=1}^{m_b}
  \pi_{i,j} =w_{i}^{( a )} ,\;  \forall i\in\mathcal{I}_a\;. \label{eq:primal} 
\end{array}
\end{equation}

We call $\{\pi_{i,j}\}$ the {\em matching weights} between support points
$x_i^{(a)}$ and $x_j^{(b)}$ or the {\em optimal coupling} for $P^{(a)}$ and $P^{(b)}$. In D2-clustering, we use the $L_2$ Wasserstein distance. From now on, we will denote $W_2$ simply by $W$.

\begin{algorithm}[ht!]
\caption{D2 Clustering~\cite{li2008real}}\label{alg:clustering}
\begin{algorithmic}[1]
\Procedure{D2Clustering}{$\{ P^{( k )} \}_{k=1}^{M}$, $K$}
  \State Denote the label of each objects by $l^{( k )}$.
  \State Initialize $K$ random centroid $\{ Q^{( i )} \}_{i=1}^{K}$.
  \Repeat
  \For{$k=1, \ldots ,M$}
  \Comment{Assignment Step}
  \State $l^{( k )} \assign \tmop{argmin}_{i}  W ( Q^{( i )},P^{( k )} )$;
  %\State store $\{ \pi_{i,j}^{( k )} \}_{k=1}^{M}$ subject to
  %  $W ( Q^{( l^{( k )} )} ,P^{( k )} )$;
  \EndFor    
  \For{$i=1, \ldots ,K$}
 \Comment{Update Step}    
 \State$Q^{( i )} \assign \tmop{argmin}_{Q} \sum_{l^{( k )}
    =i} W ( Q,P^{( k )} )$ (*)
 \EndFor    
  \Until{the number of changes of $\{ l^{( k )} \}$ meets some stopping criterion}
  
  \State \Return $\{ l^{( k )} \}_{k=1}^{M}$ and $\{ Q^{( i )} \}_{i=1}^{K}$.
  \EndProcedure
\end{algorithmic}
\end{algorithm}

%\subsection{D2-Clustering}\label{sec:d2c}
Consider a set of discrete distributions $\{P^{(k)}, k=1, ..., \bar{N}\}$,
where $P^{(k)}=\{(w_i^{(k)}, x_i^{(k)}), i=1, ..., m_k\}$.  The goal of
D2-clustering is to find a set of centroid distributions $\{Q^{(i)}, i=1,
..., K\}$ such that the total within-cluster variation is minimized:
\[\displaystyle
\min_{\{Q^{(i)}\}}\sum_{k=1}^{\bar{N}}\min_{i=1, ..., K}W^2(Q^{(i)}, P^{(k)})\;
.\]
Similarly as in K-means, D2-clustering alternates the optimization of the
centroids $\{Q^{(i)}\}$ and the assignment of each instance to the nearest centroid, the iteration referred to as the {\em outer loop} (Algorithm~\ref{alg:clustering}). 
The major computational challenge in the algorithm is to compute the optimal
centroid for each cluster at each iteration.  This computation also marks the main difference between
D2-clustering and K-means in which the optimal centroid is in a simple closed
form.  The new scalable algorithms we develop here aim primarily to expedite this optimization step.  
For clarity of presentation, we now focus 
on this optimization problem and describe the notation below.
Suppose we have a set of discrete distributions
$\{ P^{( 1 )} , \ldots ,P^{( N )} \}$. $N$ is the sample size for computing one Wasserstein barycenter. We want to find a centroid $P^{} :  \{
( w_{1} ,x_{1} ) , \ldots , ( w_{m} ,x_{m} ) \}$, such that
     \begin{equation}
  \min_{P} \frac{1}{N} \sum_{k=1}^{N} W^2 ( P,P^{( k )} ) \label{eq:centroid}
\end{equation}
with respect to the weights and support points of $P$. This is the
\textbf{main question} we tackle in this paper.  There is an implicit layer of
optimization in {\eqref{eq:centroid}}---the computation of $W^2 (
P,P^{( k )} )$.  The variables in optimization {\eqref{eq:centroid}} thus
include the weights in the centroid $\{ w_{i} \in \mathbbm{R}^{+} \}$, the
support points $\{ x_{i} \in \mathbbm{R}^{d} \}$, and the optimal coupling
between $P$ and $P^{( k )}$ for each $k$, denoted by $\{ \pi_{i,j}^{( k )} \}$ (see
Eq. {\eqref{eq:primal}}).

To solve {\eqref{eq:centroid}}, D2-clustering alternates the optimization of 
$\{ w_{i} \}$ and $\{ \pi_{i,j}^{( k )}
\}$, $k=1, ..., N$, versus $\{ x_{i} \}$.  
\begin{enumerate}
\item
$\Delta_k$ denotes a probability simplex of $k$ dimensions.  
\item 
$\mathbf{1}$ denotes a vector with all elements equal to one.
\item
$\tmmathbf{x} = ( x_{1} , \ldots ,x_{m} )
\in \mathbbm{R}_{d \times m}$, 
$\tmmathbf{w} = ( w_{1} , \ldots ,w_{m} ) \in \Delta_{m}$.  
\item
$\tmmathbf{x}^{(k)} = ( x^{(k)}_{1} , \ldots ,x^{(k)}_{m_k} )
\in \mathbbm{R}_{d \times m_k}$, $k=1, ..., N$. 
\item
$\tmmathbf{w}^{(k)} = (w_1^{(k)},\ldots,w_{m_k}^{(k)})\in \Delta_{m_k}$\ .
\item 
$C(\tmmathbf{x}, \tmmathbf{x}^{(k)} ) = (\|x_i - x_j^{(k)}\|^2)_{i,j}\in \mathbb{R}_{m\times m_k}$\ .
\item
$\tmmathbf{X}= ( \tmmathbf{x}^{( 1 )} , \ldots , \tmmathbf{x}^{( N
)} ) \in \mathbbm{R}_{d \times n}$, where $n=\sum\limits_{k=1}^{N} m_k$\ .
\item
$\Pi^{(k)} = ( \pi_{i,j}^{(k)} ) \in \mathbbm{R}_{m\times m_k}^{+}$, $k=1, ..., N$.  
\item
$\Pi = ( {\Pi}^{( 1 )},\ldots , {\Pi}^{(N)} ) \in \mathbbm{R}^+_{m \times n}$\ . 
\item
Index set $\mathcal{I}^{c}=\{1, ..., N\}$, $\mathcal{I}_k=\{1, ...,
m_k\}$, for\break $k\in \mathcal{I}^{c}$, and $\mathcal{I}'=\{1, ..., m\}$. 
\end{enumerate}

With $\tmmathbf{w}$ and $\Pi$ fixed, the
cost function {\eqref{eq:centroid}} 
is quadratic in terms of $\tmmathbf{x}$,
and the optimal $\tmmathbf{x}$ is solved by:
\begin{equation}
  x_{i} \assign \frac{1}{N w_{i}} \sum_{k=1}^{N} \sum_{j=1}^{m_k}
  \pi_{i,j}^{( k )}  x_{j}^{( k )} , \;\; i\in\mathcal{I}' \, ,\label{eq:updatex}
\end{equation}
Or, we can write it in matrix form: $\tmmathbf{x} \assign \frac{1}{N} X \Pi^{T}
\tmop{diag} ( 1./\tmmathbf{w} )$. 
However, with fixed $\tmmathbf{x}$, updating $\tmmathbf{w}$ and $\Pi$ is
challenging.  D2-clustering solves a large LP 
as follows:
\begin{eqnarray}
&\min\limits_{\Pi \in \mathbbm{R}^+_{m \times n}, \tmmathbf{w}\in \Delta_m}  \sum\limits_{k=1}^{N} \langle C(\tmmathbf{x},\tmmathbf{x}^{(k)}), \Pi^{(k)}\rangle\;, \label{eq:fullbatch-lp}\\
 \textrm{s.t.} & \mathbf{1}\cdot (\Pi^{(k)})^T  =\tmmathbf{w}\;,   \;\;
  \mathbf{1}\cdot \Pi^{(k)} =\tmmathbf{w}^{(k)}, \forall k\in\mathcal{I}^c\;, &  \nonumber
\end{eqnarray}
where the inner product $\langle A, B\rangle \assign \mbox{tr}(A B^t)$.

\begin{algorithm}[ht!]
\caption{Centroid Update with Full-batch LP~\cite{li2008real,cuturi2014fast}}\label{alg:centroid0}
\begin{algorithmic}[1]
\Procedure{Centroid}{$\{ P^{( k )} \}_{k=1}^{N}$}
  \Repeat
  \State Updates $\{ x_{i} \}$ from Eq.~\eqref{eq:updatex};
  \State Updates $\{ w_{i} \}$ from solving full-batch LP
  {\eqref{eq:fullbatch-lp}};
  \Until{$P$ converges}
  \State \Return $P$
\EndProcedure
\end{algorithmic}
\end{algorithm}

By iteratively solving {\eqref{eq:updatex}} and {\eqref{eq:fullbatch-lp}}, 
referred to as the {\em inner loop},
the step of updating the cluster centroid in Algorithm~\ref{alg:clustering} 
is fulfilled~\cite{li2008real,cuturi2014fast}.  
We present the centroid update step in Algorithm~\ref{alg:centroid0}.
In summary, D2-clustering is given by Algorithm~\ref{alg:clustering} with
Algorithm~\ref{alg:centroid0} embedded as one key step.  

The major difficulty in solving {\eqref{eq:fullbatch-lp}} is that a standard
LP solver typically has a polynomial complexity in terms of
the number of variables $m + \sum_{k=1}^N m_k m$, prohibiting its
scalability to a large number of discrete distributions in one cluster.  When
the cluster size is small or moderate, say dozens, it is shown
that the standard LP solver can be faster than a scalable
algorithm~\cite{ye2014scaling}.  However, when the cluster size grows, the
standard solver slows down quickly.  This issue has been demonstrated by
multiple empirical studies~\cite{li2008real,ye2014scaling,zhang2015parallel}.  

Our key observation is that in the update of a centroid distribution, the
variables in $\tmmathbf{w}$ are much more important than are the matching weights
in $\Pi$ needed for computing the Wasserstein distances.  The parameter
$\tmmathbf{w}$ is actually part of the output centroid, while $\Pi$ is not,
albeit accounting for the vast majority of the variables in
{\eqref{eq:fullbatch-lp}}.  We also note that the solution
{\eqref{eq:fullbatch-lp}} is not the end result, but rather it is one round of centroid update in the outer
loop. It is thus adequate to have a sufficiently
accurate solution to {\eqref{eq:fullbatch-lp}}, motivating us to pursue
scalable methods such as ADMM, known to be fast for reaching the vicinity of
the optimal solution.

%%%%%%%%%%%%%%%%%%%%%%%%%%
\section{Scalable Centroid Computation}\label{sec:centroid}
We propose algorithms scalable with large-scale datasets, 
and compare their performance in terms of speed and memory.
They are (a) subgradient descent with $N$ mini-LP following 
similar ideas of~\cite{cuturi2014fast} (included in Appendix~\ref{sec:descent}), 
(b) standard ADMM with $N$ mini-QP, and (c) modified B-ADMM with closed 
forms in each iteration of the inner loop.  
The bottleneck in the computation of D2-clustering is the inner loop,
detailed in Algorithm \ref{alg:centroid0}.  The approaches we develop
here all aim for fast solutions for the inner loop, that is, to improve
Algorithm \ref{alg:centroid0}.  These new methods can reduce the computation
for centroid update to a comparable (or even lower) level as the label
assignment step, usually negligible in the original D2-clustering.  
As a result, we also take measures to expedite the labeling step, with details provided
in Section \ref{sec:impl}.

\subsection{Alternating Direction Method of Multipliers}\label{sec:admm}
ADMM typically solves a problem with two sets of variables (in our case,
they are $\Pi$ and $\tmmathbf{w}$), which are only coupled in constraints,
while the objective function is separable across this splitting of the two sets (in our
case, $\tmmathbf{w}$ is not present in the objective function)~\cite{boyd2011distributed}. 
Because problem {\eqref{eq:fullbatch-lp}} has multiple sets of constraints
including both equalities and inequalities, it is not a typical scenario to apply 
ADMM. We propose to relax all equality constraints 
$\sum_{l=1}^{m_k}\pi_{i,l}^{(k)}=w_{i}$, $\forall k\in\mathcal{I}^c$, $i\in\mathcal{I}'$ in {\eqref{eq:fullbatch-lp}} to their
corresponding augmented Lagrangians and 
use the other constraints to determine a convex set for the parameters being optimized. 
Let $\Lambda=(\lambda_{i,k})$, $i\in\mathcal{I}'$, $k\in\mathcal{I}^c$.
Let $\rho$ be a parameter to balance the objective function and the augmented
Lagrangians. Define
$\Delta_{\Pi} = \left\{ ( \pi_{i,j}^{( k )} ) : \sum_{i=1}^{m}
\pi_{i,j}^{( k )} =w_{j}^{( k )} , \pi_{i,j}^{( k )} \geqslant 0, k\in\mathcal{I}^c, i\in\mathcal{I}', j\in\mathcal{I}_k
\right\}$. Recall that $\Delta_m = \left\{ (
w_{1} , \ldots ,w_{m} ) | \nobracket \sum_{i=1}^{m} w_{i} =1,w_{i} \geqslant 0
\right\}$.
As in the method of multipliers, we form the
scaled augmented Lagrangian $L \rho ( \Pi , \tmmathbf{w} , \Lambda )$ as follows
\begin{multline}
  L_{\rho} ( \Pi , \tmmathbf{w} , \Lambda ) = \sum_{k=1}^{N} \langle C(\tmmathbf{x},\tmmathbf{x}^{(k)}), \Pi^{(k)}\rangle + \\ \rho 
  \sum_{\substack{i\in\mathcal{I}'\\k\in\mathcal{I}^c} } \lambda_{i,k}
  \left( \sum_{j=1}^{m_k} \pi_{i,j}^{( k )} -w_{i} \right) +
  \frac{\rho}{2} \sum_{\substack{i\in\mathcal{I}'\\k\in\mathcal{I}^c} } \left( \sum_{j=1}^{m_k} \pi_{i,j}^{( k )}
  -w_{i} \right)^{2} \, .
\end{multline}
Problem
{\eqref{eq:fullbatch-lp}} can be solved using ADMM iteratively as follows.
\begin{eqnarray}
  \Pi^{n+1} \assign \underset{\Pi \in \Delta_{\Pi}}{\tmop{argmin}} 
  L_{\rho} ( \Pi , \tmmathbf{w}^{n} , \Lambda^{n} ) \;, \qquad \quad \label{eq:admm0}\\
  \tmmathbf{w}^{n+1}  \assign \underset{\tmmathbf{w} \in
  \Delta_m}{\tmop{argmin}}  L_{\rho} ( \Pi^{n+1} , \tmmathbf{w} ,
  \Lambda^{n} ) \;, \qquad \;\; \label{eq:admm1}\\
  \lambda_{i,k}^{n+1}  \assign \lambda_{i,k}^{n} + \sum_{j=1}^{m_k}
  \pi_{i,j}^{( k ) ,n+1} -w_{i}^{n+1} \, , i\in\mathcal{I}', \, k\in\mathcal{I}^c \,. \label{eq:admm2}
\end{eqnarray}
Based on {\eqref{eq:admm0}}, $\Pi$ can be updated by updating $\Pi^{(k)}$, $k=1, ..., N$ separately.  Comparing with the full batch LP in {\eqref{eq:fullbatch-lp}} which solves all $\Pi^{(k)}$, $k=1, ..., N$, together, ADMM solves instead $N$ disjoint constrained quadratic programming (QP). This step is the key for achieving computational complexity linear in $N$, the main motivation for employing ADMM. Specifically, we solve {\eqref{eq:fullbatch-lp}} by solving {\eqref{eq:admmqp}} below for each $k=1, ..., N$:
\begin{equation}
\begin{array}{rl}
  \min_{\pi_{i,j}^{( k )} \geqslant 0} & \langle C(\tmmathbf{x},\tmmathbf{x}^{(k)}), \Pi^{(k)}\rangle \\ & +
   \dfrac{\rho}{2} \sum_{i=1}^{m} \left( \sum_{j=1}^{m_k} \pi_{i,j}^{( k )} -w_{i}^{n} + \lambda_{i,k}^{n} \right)^{2} \\
  \textrm{s.t.} & \mathbf{1}\cdot \Pi^{( k )} =\tmmathbf{w}^{( k )}, k\in\mathcal{I}^c.
\end{array}\label{eq:admmqp}
\end{equation}
Since we need to solve small-size problem {\eqref{eq:admmqp}} in multiple
rounds, we prefer active set method with warm start. Define 
$\tilde{w}_{i}^{(k),n+1} \assign \sum_{j=1}^{m_k} \pi_{i,j}^{( k )
,n+1} + \lambda_{i,k}^{n}$ for $i=1, ..., m$, $k=1, ..., N$. We can rewrite
step {\eqref{eq:admm1}} as
\begin{equation}
  \min_{\tmmathbf{w}\in \Delta_m} \sum_{i=1}^{m}\sum_{k=1}^{N} ( \tilde{w}_{i}^{(k),n+1} -w_{i} )^{2} \;.
 % \nonumber
 \label{eq:admmw}
\end{equation}
\begin{algorithm}[ht!]
\caption{Centroid Update with ADMM~\cite{ye2014scaling}}\label{alg:admmqp}
\begin{algorithmic}[1]
\Procedure{Centroid}{$\{ P^{( k )} \}_{k=1}^{N}$, $P$, $\Pi$}

\State Initialize $\Lambda^{0} =0$ and $\Pi^{0} \assign \Pi$.

\Repeat

\State Updates $\{ x_{i} \}$ from Eq.{\eqref{eq:updatex}};

\State Reset dual coordinates $\Lambda$ to zero;

\For {$iter = 1, \ldots, T_{admm}$}
\For{$k=1, \ldots ,N$}

\State Update $\{ \pi_{i,j} \}^{( k )}$ based on QP
Eq.{\eqref{eq:admmqp}};
\EndFor
\State Update $\{ w_{i} \}$ based on QP Eq.{\eqref{eq:admmw}};
\State Update $\Lambda$ based on Eq. \eqref{eq:admm2};
\EndFor
 \Until $P$ converges
  \State \Return $P$
\EndProcedure
\end{algorithmic}
\end{algorithm}

We summarize the computation of the centroid distribution $P$ for distributions $P^{(k)}$, $k=1, ..., N$, in
Algorithm~\ref{alg:admmqp}. There are two hyper-parameters to choose: $\rho$ and the number of iterations $T_{admm}$.
We empirically select $\rho$ proportional to the averaged transportation costs: 
\begin{equation}
\rho = \dfrac{\rho_0}{N n m} \sum_{k=1}^N\sum_{i\in \mathcal{I}'} \sum_{j\in \mathcal{I}_k} c(x_i, x_j^{(k)}) \;.
\label{eq:rho}
\end{equation}

Let us compare the computational efficiency of ADMM and the subgradient descent method.  
In gradient descent based approaches, it is costly to choose an effective step-size along the descending 
direction because at each search point, we need to solve $N$ LP --- an issue also discussed in~\cite{benamou2014iterative}.  
ADMM solves $N$ QP sub-problems instead of LP.  
The amount of computation in each sub-problem of ADMM is thus 
usually higher and grows faster with the number of support 
points in $P^{(k)}$'s.  Whether 
the increased complexity at each iteration of ADMM 
is paid off by a better convergence rate ({\it i.e.}, a smaller number of iterations) is unclear.  
The computational limitation of ADMM caused by QP motivates us 
to explore B-ADMM that avoids QP in each iteration.

\subsection{Bregman ADMM}\label{sec:badmm}
Bregman ADMM (B-ADMM) replaces the quadratic augmented Lagrangians by the Bregman
divergence when updating the split variables~\cite{bregman1967relaxation}. 
Similar ideas trace back at least to the
early 1990s~\cite{censor1992proximal,eckstein1993nonlinear}.
We adapt the design in \cite{wang2014bregman,cuturi2013sinkhorn} for
solving the OT problem with a large
set of support points. 
Consider two sets of variables $\Pi_{(k,1)}=(\pi_{i,j}^{( k,1 )})$, 
$i\in\mathcal{I}'$, $j\in\mathcal{I}_k$, and
$\Pi_{(k,2)}=(\pi_{i,j}^{( k,2 )})$, $i\in\mathcal{I}'$, $j\in\mathcal{I}_k$, for $k=1, ..., N$ under the following constraints. Let
\begin{eqnarray}
  \Delta_{k,1} :=  \left\{ \pi_{i,j}^{(k,1)} \geqslant 0:
   \sum_{i=1}^{m} \pi_{i,j}^{(k,1)} =w_{j}^{( k )} ,j\in\mathcal{I}_k \right\}\;,\\
  \Delta_{k,2} ( \tmmathbf{w} ) :=  \left\{ \pi_{i,j}^{(k,2)}
  \geqslant 0 :\sum_{j=1}^{m_k} \pi_{i,j}^{(k,2)} =w_{i}
   ,i\in\mathcal{I}' \right\},
\end{eqnarray}
then
$\Pi^{( k,1 )} \in \Delta_{k,1}$ and $\Pi^{( k,2 )} \in \Delta_{k,2} ( \tmmathbf{w} )$.
We introduce some extra notations: 
\begin{enumerate}
\item $\bar{\Pi}^{(1)}=\{\Pi^{(1,1)}, \Pi^{(2,1)}, \ldots, \Pi^{(N,1)}\}$,
\item $\bar{\Pi}^{(2)}=\{\Pi^{(1,2)}, \Pi^{(2,2)}, \ldots, \Pi^{(N,2)}\}$, 
\item $\bar{\Pi}=\{\bar{\Pi}^{(1)}, \bar{\Pi}^{(2)}\}$,
\item $\Lambda =\{\Lambda^{(1)},\ldots,\Lambda^{(N)}\}$, where 
$\Lambda^{(k)}=(\lambda_{i,j}^{(k)})$, $i\in\mathcal{I}'$, $j\in\mathcal{I}_k$, 
is a $m\times m_k$ matrix.
\end{enumerate} 

B-ADMM solves {\eqref{eq:fullbatch-lp}} by treating the augmented
Lagrangians conceptually as a designed divergence between $\Pi^{(k,1)}$ and $\Pi^{(k,2)}$, adapting to the updated variables. It restructures the original problem \eqref{eq:fullbatch-lp} as
\begin{eqnarray}
\underset{\bar{\Pi},\tmmathbf{w}}{\min} && \sum_{k=1}^{N} \langle C(\tmmathbf{x},\tmmathbf{x}^{(k)}), \Pi^{(k,1)}\rangle \;,\\
\textrm{s.t.} && \tmmathbf w \in \Delta_m \;, \nonumber \\
&& \Pi^{(k,1)}\in \Delta_{k,1}, \quad \Pi^{(k,2)}\in \Delta_{k,2}(\tmmathbf w),\;\; k=1,\ldots,N \;,\nonumber \\ 
&&  \Pi^{(k,1)} = \Pi^{(k,2)}, \;\; k=1,\ldots,N\;. \nonumber
\label{eq:badmm_prob}
\end{eqnarray}
Denote the dual variables $\Lambda^{(k)}=(\lambda_{i,j}^{(k)})$, $i\in\mathcal{I}'$, $j\in\mathcal{I}_k$, for $k=1, ..., N$. 
Use $\tmop{KL} ( \cdot , \cdot )$ to denote the Kullback--Leibler divergence between two distributions.
The B-ADMM algorithm adds the augmented Lagrangians for the last
set of constraints in its updates, yielding the following equations.
\begin{eqnarray}
&&\!\!\!\!\!\!\!\!\bar{\Pi}^{( 1 ) ,n+1} \assign \underset{\{\Pi^{( k,1 )} \in
  \Delta_{k,1}\}}{\tmop{argmin}} \sum_{k=1}^{N} \Bigg( \langle C(\tmmathbf{x},\tmmathbf{x}^{(k)}), \Pi^{(k,1)}\rangle \nonumber\\
  && \;\;\;\;\;+ 
  \langle \Lambda^{( k ) ,n}, \Pi^{( k,1 )}\rangle + 
  \rho \tmop{KL} ( \Pi^{( k,1 )} , \Pi^{( k,2 ) ,n} ) \Bigg),
  \label{eq:badmm0}\\
&&\!\!\!\!\!\!\!\!\bar{\Pi}^{( 2 ) ,n+1} , \tmmathbf{w}^{n+1} \assign
  \underset{\substack{\{\Pi^{( k,2 )}
  \in \Delta_{k,1} ( \tmmathbf{w} )\} \\ \tmmathbf{w} \in
  \Delta_m} }{\tmop{argmin}} \sum_{k=1}^{N} \Bigg( - \langle \Lambda^{( k ) ,n} 
  , \Pi^{( k,2 )} \rangle \nonumber \\
   && \;\;\;\;\;+ \rho \tmop{KL} ( \Pi^{( k,2 )} , \Pi^{( k,1 ) ,n+1} ) \Bigg),\label{eq:badmm1}\\
&&\!\!\!\!\!\!\!\!\Lambda^{ n+1} \assign \Lambda^{ n} + \rho ( \bar{\Pi}^{( 1 ) ,n+1} - \bar{\Pi}^{(2 ) ,n+1} ).  \label{eq:badmm2}
\end{eqnarray}
We note that if $\tmmathbf{w}$ is fixed,
\eqref{eq:badmm0} and \eqref{eq:badmm1}
can be split by index $k=1, ..., N$, and have closed form solutions for each $k$. Let $eps$ be the floating-point tolerance ({\it e.g.} $10^{-16}$).
For any $i\in \mathcal{I}', \; j\in\mathcal{I}_k$,
\begin{eqnarray}
 &&\!\!\!\!\!\!\!\!\!\!\!\!\!\! \tilde{\pi}_{i,j}^{( k,2 ) ,n} \assign \pi_{i,j}^{( k,2 ) ,n} \exp\!\!
  \left[ \frac{c \left( x_{i} ,x_{j}^{( k )}\! \right) \!+\! \lambda_{i,j}^{(k), n}}{-\rho}
  \right]\!\!+\!eps \;,\label{eq:badmmc0b} \\
 &&\!\!\!\!\!\!\!\!\!\!\!\!\!\! \pi_{i,j}^{( k,1 ) ,n+1} \assign  \dfrac{\tilde{\pi}_{i,j}^{( k,2
  ) ,n}}{{\sum_{l=1}^{m}} \tilde{\pi}_{l,j}^{( k,2 ) ,n}}\cdot 
  w_{j}^{( k )} \;,\label{eq:badmmc0}  \\
 &&\!\!\!\!\!\!\!\!\!\!\!\!\!\! \tilde{\pi}_{i,j}^{( k,1 ) ,n+1} \assign  \pi_{i,j}^{( k,1 ) ,n+1} \exp \left[
  \frac{1}{\rho} \lambda_{i,j}^{(k), n} \right]\!+\!eps \;,  \label{eq:badmmc1b} \\
 &&\!\!\!\!\!\!\!\!\!\!\!\!\!\! \pi_{i,j}^{( k,2 ) ,n+1} \assign  \dfrac{\tilde{\pi}_{i,j}^{( k,1 )
  ,n+1}}{{\sum_{l=1}^{m_k}} \tilde{\pi}_{i,l}^{( k,1 ) ,n+1}}\cdot 
  w_{i} \; .  \label{eq:badmmc1} 
\end{eqnarray}
Because we need to update $\tmmathbf{w}$ in each iteration, it is not easy to solve
\eqref{eq:badmm1}.  We consider decomposing \eqref{eq:badmm1} into two stages.
Observe that the minimum value of \eqref{eq:badmm1} under a given
$\tmmathbf{w}$ is 
\begin{equation}
  \min_{\tmmathbf w\in \Delta_m} \sum_{k=1}^{N} \sum_{i=1}^{m} 
  w_{i} \left[ \log ( w_{i} ) - \log \left(
  \sum\nolimits_{j=1}^{m_k} \tilde{\pi}_{i,j}^{( k,1 ) ,n+1} \right) \right].
  \label{eq:consensus0}
\end{equation}
The above term (a.k.a. the consensus operator) is minimized by
\begin{equation}
  w_{i}^{n+1} \propto \left[ \prod\limits_{k=1}^{N} \left( \sum\limits_{j=1}^{m_k} \tilde{\pi}_{i,j}^{( k,1 ) ,n+1} \right) \right]^{{1}/{N}},\;\;
  \sum\limits_{i=1}^{m} w_{i}^{n+1} = 1\;.
\end{equation}
However, the above equation is a geometric mean, which is numerically
unstable when $\sum_{j=1}^{m_k} \tilde{\pi}_{i,j}^{( k,1 ) ,n+1}
\rightarrow 0^{+}$ for some combination of $i$ and $k$.  
%Similar numerical problem also arises in the alternating
%Bregman projection algorithm~\cite{benamou2014iterative}, where a
%regularization scheme was introduced to smooth the objective function.  
Here, we employ a different technique.  Let 
\[
\tilde{w}_{i}^{( k,1 ) ,n+1} \propto
\sum_{j=1}^{m_k} \tilde{\pi}_{i,j}^{( k,1 ) ,n+1}, \;\; \textrm{s.t.} 
\sum_{i=1}^{m}
\tilde{w}_{i}^{( k,1 ) ,n+1} =1.
\]
Let the distribution
$\tilde{\tmmathbf{w}}^{(k),n+1}= (\tilde{w}_{i}^{( k,1 ) ,n+1})_{i=1, ..., m} $.
Then Eq. {\eqref{eq:consensus0}} is equivalent to $\min_{\tmmathbf w\in
  \Delta_m} \sum_{k=1}^{N}
\mbox{KL}({\tmmathbf{w}},\tilde{\tmmathbf{w}}^{(k),n+1})$.  Essentially, a consensus
${\tmmathbf{w}}$ is sought to minimize the sum of KL divergence.  In the same
spirit, we propose to find a consensus by changing the order of
${\tmmathbf{w}}$ and $\tilde{\tmmathbf{w}}^{(k),n+1}$ in the KL divergence:
$ \min_{\tmmathbf w\in \Delta_m}\sum\limits_{k=1}^{N}
      \tmop{KL}(\tilde{\tmmathbf{w}}^{(k),n+1}, {\tmmathbf{w}})= $
\begin{equation}
  \min_{\tmmathbf w\in \Delta_m}
 \sum\limits_{k=1}^{N}  \sum\limits_{i=1}^{m} \tilde{w}_{i}^{( k,1 ) ,n+1} (  \log ( \tilde{w}_{i}^{( k,1 ) ,n+1} ) - \log ( w_{i} ) )\; ,
  \label{eq:consensus1}
\end{equation}
which again has a closed form solution:
\begin{equation}
 \mbox{(R1)}: w_{i}^{n+1} \propto \frac 1N \sum\limits_{k=1}^{N} \tilde{w}_{i}^{( k,1 ) ,n+1} ,\;\;
  \sum\limits_{i=1}^{m} w_{i}^{n+1} = 1\;.
  \label{eq:badmmw}
\end{equation}
The solution of Eq.~\eqref{eq:consensus1} overcomes the numerical
instability. We will call this heuristic update rule as (R1),
which has been employed in the Bregman clustering method~\cite{banerjee2005clustering}.
In addition, a slightly different version of update rule can be
\begin{equation}
 \mbox{(R2)}: \left(w_{i}^{n+1} \right)^{1/2} 
 \propto \frac 1N \sum\limits_{k=1}^{N} \left(\tilde{w}_{i}^{( k,1 ) ,n+1}\right)^{1/2} ,\;\;
  \sum\limits_{i=1}^{m} w_{i}^{n+1} = 1\;.
  \label{eq:badmmw2}
\end{equation}
In Section~\ref{sec:compare}, we conduct experiments for testing both (R1) and (R2).
We have tried other update rules, such as Fisher-Rao Riemannian center~\cite{srivastava2007riemannian}, and found that the 
experimental results do not differ much in terms of the converged objective function.
It is worth mentioning that neither (R1) nor (R2) ensures the 
convergence to a (local) minimum. 

We summarize the B-ADMM approach in Algorithm~\ref{alg:badmm}. The
implementation involves one hyper-parameters $\rho$ (by default, $\tau=10$). 
In our implementation, we choose $\rho$ relatively according to Eq.~\eqref{eq:rho}.
To the best of our knowledge, the convergence of B-ADMM has not been
proved for our formulation (even under fixed support points $\tmmathbf{x}$) 
although this topic has been 
pursued in recent literature~\cite{wang2014bregman}. In the general case of solving
Eq.~\eqref{eq:centroid}, the optimization of the cluster centroid is
non-convex because the support points are updated after B-ADMM is
applied to optimize the weights. In Section \ref{sec:convergence}, we empirically
test the convergence of the centroid optimization algorithm based on 
B-ADMM.  We found that B-ADMM usually converges quickly 
to a moderate accuracy, making it preferable for D2-clustering.
In our implementation, we use a fixed number 
of B-ADMM iterations (by default, 100) across multiple assignment-update rounds in D2-clustering.
\begin{algorithm}[htp]
\caption{Centroid Update with B-ADMM}
\label{alg:badmm}
\begin{algorithmic}[1]
\Procedure{Centroid}{$\{ P^{( k )} \}_{k=1}^{N}$, $P^{}$, $\Pi$}.
  
  \State $\Lambda \assign 0$; $\bar{\Pi}^{( 2 ) ,0} \assign \Pi$.
  
  \Repeat
  
  \State Update $\tmmathbf{x}$ from Eq.{\eqref{eq:updatex}} per $\tau$ loops;
  \For{$k=1, \ldots ,N$}
  \State Update $\Pi^{( k,1 )}$ based on Eq.{\eqref{eq:badmmc0b} \eqref{eq:badmmc0}};
  \State Update $\{\tilde{\pi}_{i,j}^{( k,1 )}\}$ based on Eq.{\eqref{eq:badmmc1b}}; 
  \EndFor
  \State Update $\tmmathbf{w}$ based on Eq.{\eqref{eq:badmmw}} or Eq.{\eqref{eq:badmmw2}};
  \For{$k=1,\ldots, N$}
   \State Update $\Pi^{( k,2 )}$ based on Eq.{\eqref{eq:badmmc1}};
   \State $\Lambda^{( k )} \assign \Lambda^{( k )} + \rho ( \Pi^{( k,1)} - \Pi^{( k,2 )} )$;
  \EndFor  
  
  \Until $P$ converges
  \State \Return $P$
\EndProcedure
\end{algorithmic}
\end{algorithm}

\section{Comparison between B-ADMM and Iterative Bregman Projection}~\label{sec:compare}
Both the B-ADMM and IBP can be rephrased into two-step 
iterative algorithms via mirror maps (in a similar way of 
mirror prox~\cite{nemirovski2004prox} or mirror descent~\cite{beck2003mirror}).
One step is the free-space move in the dual space, and
the other is the Bregman projection (as used in IPFP) in the primal space. 
Let $\Phi(\cdot)$ be the entropy function,
the mirror map used by IBP is $\Phi(\Pi)$, while the mirror map of B-ADMM is $$\Phi(\Pi^{(1)},\Pi^{(2)},\Lambda)=\Phi(\Pi^{(1)}) + \Phi(\Pi^{(2)}) 
+ \dfrac{\|\Lambda\|^2}{\rho^2},$$ where $\Lambda$ is the dual coordinate derived from
relaxing constraints $\Pi^{(1)}=\Pi^{(2)}$ to a saddle point reformulation. 
IBP alternates the move 
$-\left[\dfrac{C}{\varepsilon}\right]$ in the dual space and projection in $\Delta_1$ or $\Delta_2$ of the primal space.
In comparison, B-ADMM alternates the move
$$
-\begin{bmatrix}
\dfrac{C+\Lambda}{\rho}+\nabla\Phi(\Pi^{(1)})-\nabla\Phi(\Pi^{(2)})\\
-\dfrac{\Lambda}{\rho}+\nabla\Phi(\Pi^{(2)})-\nabla\Phi(\Pi^{(1)})\\
\rho (\Pi^{(1)}-\Pi^{(2)})
\end{bmatrix},
$$
and the projection in $\Delta_1\times\Delta_2\times \mathbb R_{m_1\times m_2}$.%
\footnote{The update is done in the Gauss-Seidel type, not in the usual Jacobi type.} 
The convergence of B-ADMM is not evident from the
conventional optimization literature~\cite{bubeck2015convex} because the move is not monotonic (It is still monotonic for standard ADMM).
We conduct two pilot studies to compare B-ADMM and IBP in terms of convergence behavior and the quality of the Wasserstein barycenters with 
a sparse finite support set, where quality is measured by the objective function achieved. 

Although the second pilot study shows certain advantages of B-ADMM, we clarify that the study is not intended to demonstrate which algorithm is better in the general context of the OT problem. In fact, for the OT problem alone, IBP is more solid in theory than B-ADMM in two aspects. First, IBP has linear convergence, while B-ADMM only has a sub-linear rate~\cite{wang2014bregman}. Second, IBP yields an OT solution more accurately satisfying the coupling constraints than B-ADMM can offer with the same computational time. In the Wasserstein barycenter problem we tackle here, either algorithm must be embedded into an outer loop, which calls for practical considerations other than solving a stand-alone OT problem. We will elaborate on these points below.

\begin{figure}[ht!]
\includegraphics[width=.5\textwidth,trim={0.6cm 2.4cm 4.95cm 2.5cm},clip]{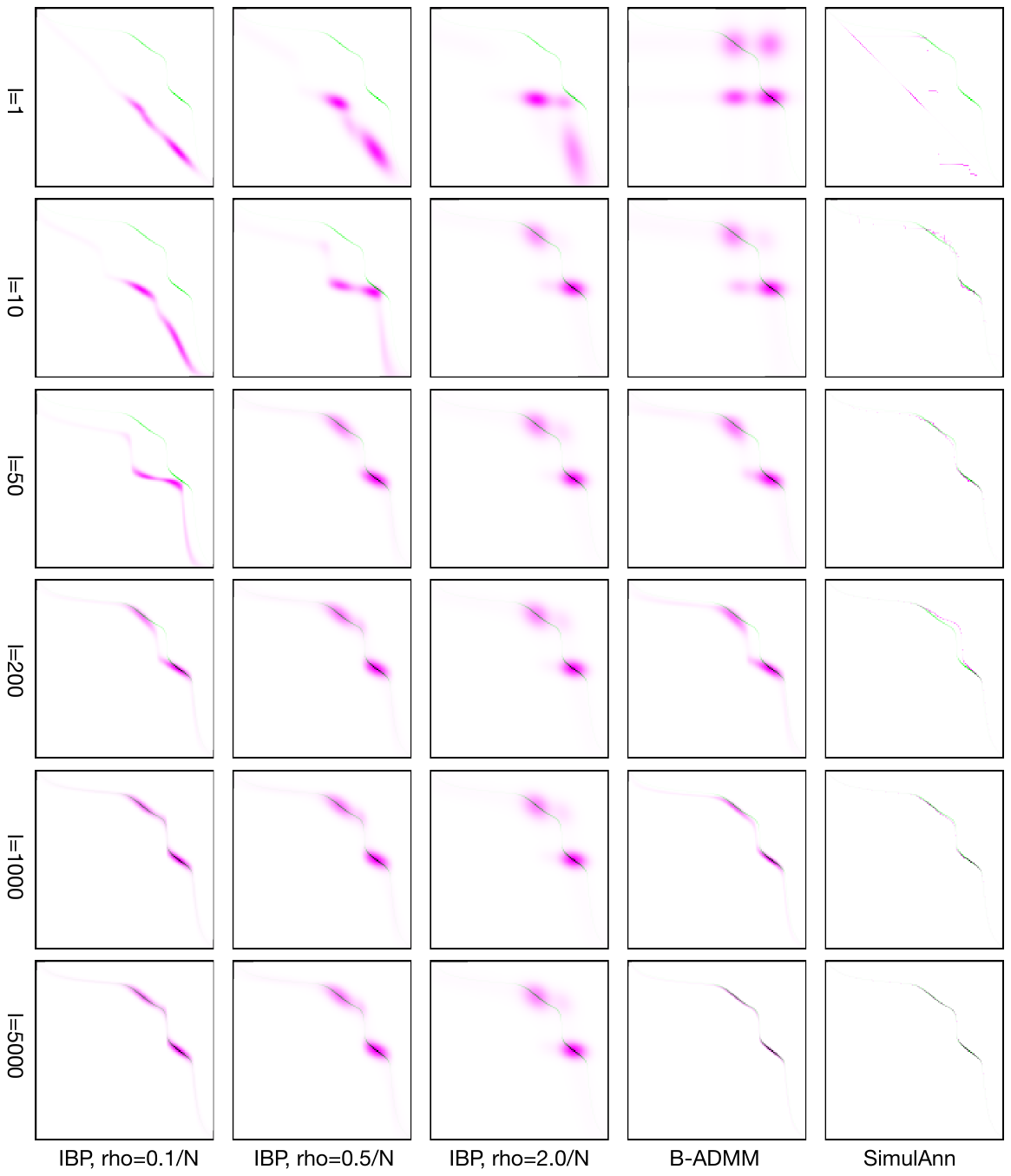}
\caption{Convergence behavior of IBP and B-ADMM. 
The solutions by IBP and B-ADMM are shown in pink, 
while the exact solution by linear programming is shown in green.
Images in the rows from top to bottom present results at different iterations 
$\{1, 10, 50, 200, 1000, 5000\}$;
The left three columns are by IBP with $\varepsilon=\{0.1/N, 0.5/N, 2/N\}$,
where $[0,1]$ is discretized with $N=128$ uniformly spaced points. The last column is by B-ADMM. This comparison is not intended for speed. IBP is about 10 to 15 times faster per iteration in this experiment. %\textcolor{red}{Please note that the comparison is solely on visualizing the behavior of pre-converged solutions, not on convergence speed. In our implementation, IBP is more than 10 times faster per iteration.}
}\label{fig:convergence}
\end{figure}

%Although the convergence has been established in~\cite{wang2014bregman}, its 
%convergence behavior for solving OT was not investigated. 
Benamou {\it et al.}~\cite{benamou2014iterative}, in their Figure 1, show how their algorithm progressively shifts mass away from the diagonal over the iterations.
We adopt the same study here and visualize in Fig.~\ref{fig:convergence} how the mass transport
between two 1D distributions evolves over iterations. 
As a qualitative study, we observe that the entropy regularization term in IBP introduces clearly smoothing effect on the final solutions. 
{It has been pointed out in the literature that the smoothing effect of entropy regularization diminishes as parameter $\varepsilon$ decreases. In our study, we found the smoothing effect is also affected by the support size of a distribution. A smooth distribution with large support tends to have higher entropy, thus a relatively smaller $\varepsilon$ is needed
to achieve similar results.}
Fig.~\ref{fig:convergence} shows that at $\varepsilon=0.1/N$, the mass transport of IBP at 5,000 iterations is close to
the unregularized solution albeit with noticeable difference. 
Setting $\varepsilon$ even smaller (say $0.04/N$) introduces double-precision overflow.
%\footnote{
As suggested by one of the reviewers, this numerical difficulty can be addressed by thresholding entries that are too small. We applied the reviewer's suggested IBP code with this technique implemented and obtained an incorrect coupling result with artifacts unexplainable by smoothing, {\it e.g.}, a non-zero region separated from the correct non-zero region. Yet more recently, we learned from the reviewer that active research on the thresholding technique is currently underway with new manuscripts emerging while we approached the final revision of this paper. In particular, log domain scaling or more sophisticated schemes have been proposed to stabilize the low $\varepsilon$ case for Sinkhorn algorithm~\cite{chizat2016scaling,schmitzer2016stabilized}. At extra computational costs, these new methods produce sharper coupling results than the standard IBP does. Investigating the effectiveness of those algorithms for the Wasserstein barycenter problem is an interesting future work.
In contrast, the output of mass transport by B-ADMM ($\rho_0=2$, the default setting) at 5,000 iteration is nearly indiscernible from the unregularized solution by LP. This example suggests that if the smoothing effect on the final coupling is to be avoided, B-ADMM may be preferred over IBP with $\epsilon\rightarrow 0$, albeit at a cost of increased computation time. With our implementation of B-ADMM and the IBP code provided by the reviewer, we found that IBP is 10 to 15 times faster per iteration.

Because B-ADMM does not require $\rho_0\!\to\!0$, no numerical difficulty has been encountered in practice. In fact, the convergence speed of B-ADMM is proven to be independent with $\rho_0$~\cite{wang2014bregman}. As for the Wasserstein barycenter problem, the minimal required tuning makes embedding B-ADMM in an outer loop easy. Putting the IBP in the outer loop reduces its speed advantage. In existing machine learning practice as well as with our algorithm here, pre-converged solutions of IBP or B-ADMM are used, making the use of a very small $\epsilon$ in IBP unnecessary. In our algorithm, however, B-ADMM per iteration slows the process as does updating barycenter support points in high dimensions. Thus, the computational gain from replacing B-ADMM by IBP is clipped by the notable proportion of time required to update the support points. In addition, although the coupling weights solved by B-ADMM do not satisfy the constraints as well as those by IBP do, they are auxiliary variables, while the barycenters are primary. 
In summary, the edge of either IBP or B-ADMM over the other subsides when they are embedded in our algorithm. 
In order to quantitatively compare the methods by measures most pertinent to the users, we examine 
the objective function and computation time in the second pilot study below.

%\textcolor{red}{In summary, if the smoothing effects of IBP solution
%is not harmful in practice, we recommend to use IBP because it is 
%always much faster per iteration.\footnote{
%\textcolor{red}{Based on our MATLAB implementation, IBP is more than 
%10 times faster than B-ADMM per iteration.}}
%and enjoys the linear convergence. When the smoothing effects of IBP solution deteriorate the accuracy
%in the outer loop, B-ADMM is a more robust supplement with increased computational cost. 
%This is the case in our developed algorithm, 
%the solved coupling is also used to update the support points of barycenter.
%In our barycenter algorithm, the B-ADMM iterations are not the only 
%bottleneck ---
%so are the updates of barycenter support points (if they are high dimensional). 
%Hence, replacing B-ADMM with IBP does not always
%significantly reduce the overall computation time by magnitudes.}
%\textcolor{red}{We emphasize that while IBP is more favorable in terms of its convergence speed,
%the existing machine learning practice of IBP still prefer to use pre-converged solution and a large $\epsilon$. In a similar spirit, we also advocate to use B-ADMM with 
%early stop, our pilot studies show that 
%their pre-converged solutions can be quite different. This poses the question which one might 
%be more effective if the pre-converged solutions are adopted to update other variables.}

For the second pilot study, we generated a set of 1,000 discrete distributions, each with a sparse finite support set obtained by clustering pixel colors of images~\cite{li2008real} ($d=3$). The average number of support 
points is around 6. Starting from the same initial estimate, 
we calculate the approximate Wasserstein barycenter by different methods. 
We obtained results for two cases: 
barycenters with support size $m=6$ 
and barycenters with support size $m=60$.
We use relatively small values of $m$ here in comparison with the existing applications of IBP in imaging science because our focus is on large data size but low support size (sparse support). 
%\textcolor{red}{This is worth noting that this setting is quite different from the usual application context of IBP in imagining science, where distributions are density maps with more than hundreds of support points.}
The obtained barycenters are then evaluated by comparing the objective function (Eq.~\eqref{eq:centroid}) with that solved directly by the full batch LP (Algorithm~\ref{alg:centroid0})---the yardstick in the comparison.

The bare form of IBP treats the case of pre-fixed support points while B-ADMM does not constrain the locations. In order to compare the two methods on a common ground, we used two tracks of experiments. In the first track the locations of support points in the barycenters are fixed and shared by all the algorithms, while in the second track both locations and weights are optimized.
To adopt IBP in the second track, we experimented with a version of IBP that can automatically restart and update its support points. The restart criterion is that the constraint satisfies certain conditions described in~\cite{cuturi2014fast,benamou2014iterative}. Generally speaking, the larger $\varepsilon_0$ is the fewer iterations are needed before a restart is evoked.
This variant of IBP does not have a descending objective.\footnote{It is because the entropic regularization term is not considered
in the update of support points.} Therefore, we chose to terminate the iterations when $\langle C, \Pi\rangle$ is detected to increase. 
The actual implementation can be found in our supplement. 
In MATLAB, we found IBP to be approximately 10 times faster than B-ADMM per iteration. 
On the other hand, being faster in one iteration does not necessarily mean
faster speed overall. We report the exact numbers of iterations and seconds before reaching a stopping criterion for both methods.

The results of the two tracks of experiments for support size $m=6, 60$ by LP, the modified B-ADMM (R1 and R2 as explained in Section \ref{sec:badmm}) and IBP, are presented in Table~\ref{table:compare}. The performance is measured by the value achieved for the objective (a.k.a. distortion) function. The results show that in both tracks, when dealing with sparse support distribution data, B-ADMM achieves lower distortion than IBP does, and furthermore, B-ADMM is quite close to LP. The gap between B-ADMM and LP is smaller than the gap between IBP and B-ADMM. On the other hand, when $\epsilon_0$ is relatively large, IBP can be much faster. There is no tuning of hyper-parameters in B-ADMM. For IBP, $\varepsilon_0$ influences the result, but not by too much as long as it is not too small. Considering the fast speed at large $\varepsilon_0$, we may favor relatively large $\varepsilon_0$ when applying IBP. For the first track experiments, the objective function obtained by IBP is considerably larger than that B-ADMM obtains. We could not push the objective function by IBP to the same level of B-ADMM by letting $\varepsilon_0\rightarrow 0$. At $\varepsilon_0=0.01, 0.005$, because double-precision overflow occurs, triggering the thresholding trick. The IBP results with thresholding triggered are marked by star ($^\ast$) in Table~\ref{table:compare}. These results are actually much worse than the others, an observation consistent with the incorrect coupling weights obtained when this trick is applied in the first pilot study. For the second track experiments, B-ADMM still achieves lower values of the objective function, but the difference from IBP is not as remarkable.

In comparison to the full LP approach, the modified B-ADMM does not yield the exact local minimum. 
This result reminds us that the modified B-ADMM is still an approximation method, and it cannot fully replace
subgradient descent based methods that minimize the objective to a true local minimum. 

\begin{table}
\caption{Comparing the solutions of the Wasserstein barycenter by LP, modified B-ADMM (our approach) and IBP. The runtime reported is based on MATLAB implementations.}~\label{table:compare}
\begin{tabular}{@{\extracolsep{\fill}}lrcc@{\extracolsep{\fill}}}
\hline
\multicolumn{4}{c}{\textbf{Solved Wasserstein barycenter with a pre-fixed support}}\\
\multicolumn{4}{c}{$(m=6) / (m=60)$}\\
method & iterations & seconds & obj. \\\hline 
full LP & NA & - & 834.1  / 709.6 \\
Our approach (R1) & 500 / 800 & 0.78 / 13.0 &  838.6 / 713.4 \\
Our approach (R2) & 400 / 700 &  0.58 / 11.8  &  835.5 / 712.3 \\
IBP[19] $\varepsilon_0=0.2$ & 150 / 40 & 0.06 / 0.11   &  978.3 / 1073.5 \\
 IBP $\varepsilon_0=0.1$ & 620 / 80 &  0.16 / 0.23  &  965.9 / 1051.6 \\
 IBP $\varepsilon_0=0.02$ & 6,640 / 760 & 1.50 / 1.49  &  957.1 / 1039.4 \\
 IBP $\varepsilon_0=0.01$ & 17,420 / 9,460 & 4.17 / 17.0  &  960.2 / 2343.8$^\ast$ \\
 IBP $\varepsilon_0=0.005$ & 36,270 / 5,110 & 8.88 / 9.83  &  1345.1$^\ast$ / 7112.2$^\ast$ \\\hline
\\ \hline
\multicolumn{4}{c}{\textbf{Solved Wasserstein barycenter with an optimized support}}\\ 
\multicolumn{4}{c}{$(m=6) / (m=60)$}\\
 method & iterations & seconds & obj. \\\hline
full LP & 20 & -  &  717.8 / 692.3 \\
Our approach (R1) & 2,000 & 2.91 / 31.1 &   723.3 / 692.6 \\
Our approach (R2) & 2,000 &   3.02 / 32.2 &  722.7 / 692.5 \\
 IBP $\varepsilon_0=0.2$ & 190 / 60 &  0.06 / 0.19 & 733.7 / 703.5\\
 IBP $\varepsilon_0=0.1$ & 730 / 130 &  0.22 / 0.31 &  734.4 / 699.5 \\
 IBP $\varepsilon_0=0.02$ & 11,860 / 1,590 &  2.67 / 2.73 & 734.9 / 705.5 \\
 IBP $\varepsilon_0=0.01$ & 33,940 / 5,130  &  7.71 / 9.01  &   734.9 / 708.3  \\
 IBP $\varepsilon_0=0.005$ & 69,860 / 16,910  &  16.5 / 30.87  &   736.1 / 708.6\\
 \hline
\end{tabular}
\end{table}

\section{Algorithm Initialization and Implementation}\label{sec:impl}
In this section, we explain some specifics in the implementation of the
algorithms, such as initialization, warm-start in optimization, measures for further speed-up, and the method for parallelization.  
%We then analyze the computation and memory complexity of the proposed algorithms.  

%In Section \ref{sec:stable},
%numerical comparison of the computational efficiency of the algorithms is
%provided based on multiple data sets.

%%%%%% Implementation details %%%%%%%%%%% 
%\subsection{Initialization and Measures for Speed-up} \label{sec:warm}
The number of support vectors in the centroid distribution, denoted by $m$,
is set to the average number of support vectors in the distributions in the
corresponding cluster.  To initialize a centroid, we select randomly a
distribution with at least $m$ support vectors from the cluster.  If the
number of support vectors in the distribution is larger than $m$, we
will merge recursively a pair of support vectors according to an optimal criterion until the support size reaches $m$, similar to the process used in linkage clustering.  Consider a
chosen distribution $P=\{(w_1, x_1), ..., (w_m, x_m)\}$.  We merge $x_i$ and
$x_j$ to $\bar{x}=(w_i x_i+w_jx_j)/\bar{w}$ , where $\bar{w}=w_i+w_j$ 
is the new weight for $\bar{x}$, 
if $(i,j)$ solves
\begin{eqnarray}
\min_{i,j}{w_iw_j\|x_i-x_j\|^2}/({w_i+w_j})\, . \label{eq:merge}
\end{eqnarray}
Let the new distribution after one merge be $P'$. It is sensible to minimize the
Wasserstein distance between $P$ and $P'$ to decide which support vectors to
merge.  We note that 
\[
W^2(P, P')\leq w_i \|x_i-\bar{x}\|^2+w_i\|x_j-\bar{x}\|^2 \, .
\]
This upper bound is obtained by the transport mapping $x_i$ and $x_j$ exclusively to $\bar{x}$ and the other support vectors to themselves.  To simplify computation, we instead minimize the upper bound, which is achieved by the $\bar{x}$ given above and by the
pair $(i,j)$ specified in Eq. (\ref{eq:merge}).

The B-ADMM method requires an initialization for $\Pi^{(k,2)}$, where
$k$ is the index for every cluster member, before starting the inner loops
(see Algorithm \ref{alg:badmm}). We use a warm-start for $\Pi^{(k,2)}$.
Specifically, for the members whose cluster labels are unchanged after the
most recent label assignment, $\Pi^{(k,2)}$ is initialized by its value
solved (and cached) in the previous round (with respect to the outer loop)
. Otherwise, we initialize $\Pi^{(k,2)}=(\pi_{i,j}^{(k,2)})$, $i=1, ..., m$,
$j=1, ..., m_k$ by
$\displaystyle \pi_{i,j}^{(k,2),0} := w_i w_j^{(k)}$. 
This scheme of initialization is also applied in the first round of 
iteration when class labels are assigned for the first
time and there exists no previous solution for this parameter.

At the relabeling step ({\it i.e.}, to assign data points to centroids after
centroids are updated), we need to compute $\bar{N} K$ Wasserstein
distances, where $\bar{N}$ is the data size and $K$ is the number of
centroids.  This part of the computation, usually negligible in the original
D2-clustering, is a sizable cost in our new algorithms.  To further boost the
scalability, we employ the technique of \cite{elkan2003using} to skip 
unnecessary distance calculation by exploiting the triangle inequality of
a metric.

In our implementation, we
use a fixed number of iterations $\epsilon_i$ for all inner loops for
simplicity.  Obtaining highly accurate result for the inner loop is not crucial 
because the partition will be changed by the outer loop. For B-ADMM, we found that setting $\epsilon_i$ to tens or a hundred
suffices.  For subgradient descent and ADMM, an even smaller $\epsilon_i$ is
sufficient, {\it e.g.}, around or below ten.  The number of iterations of the outer
loop $\epsilon_o$ is not fixed, but rather is adaptively determined when a certain
termination criterion is met. 

%Although the update of the centroids conducted by the inner loop of the
%clustering algorithms is the major computational challenge we have tackled,
%the result of the inner loop optimization is not the final outcome, but an
%input for the next round of updating cluster labels.  It is thus unnecessary
%to acquire highly accurate centroids in every iteration according to the
%optimization criterion, especially at the beginning when a significant
%portion of instances are expected to change their cluster labels in
%subsequent iterations.  Nearly optimal cluster centroids are adequate to
%result in most of the changes in cluster labels.  
%The rule-of-thumb we use is to select
%$\epsilon_i$ such that the total variation drop rate with respect to time is
%similar between the label assignment step and the centroid update step in the
%beginning iterations.

%\subsection{Serial and Parallel Implementation}
%\label{sec:serial}
With an efficient serial implementation, our algorithms can be deployed to
handle moderate scale data on a single PC.  We also implemented their
parallel versions which are scalable to a large data size and a large number
of clusters.  We use the commercial solver provided by 
Mosek,\footnote{\url{https://www.mosek.com}} which is
among the fastest LP/QP solvers available. In particular, Mosek provides
optimized simplex solver for transportation problems that fits our needs
well.  
%We found it is most efficient to create a set of LP/QP task templates
%ahead and plug in a problem to the matched template of the LP/QP tasks.
%Specifically, after computing the distance between two discrete distributions
%with $n$ and $m$ support points respectively, we cache the solved task as a
%template for future tasks sharing the same number of support points,
%a.k.a. tuple $(n,m)$.  In this way, memory allocation for those template
%tasks are done only once and used repeatedly in the lifetime of program. We
%found that when the data size is large, memory allocation for the LP/QP
%problems everytime when the solution is sought can cost considerable amount
%of time.

The algorithms we have developed here are all readily parallelizable by
adopting the Allreduce framework in MPI. In our implementation, we divide
data evenly into trunks and process each trunk at one processor.  Each trunk
of data stay at the same processor during the whole program. 
%implying that our parallelized algorithms also apply directly to a distributed computation
%platform where due to storage or communication limitations it is infeasible
%to hold all the data at a central site.  
%This is in fact an emerging challenge for big data.  
We can parallelize the algorithms simply by dividing
the data because in the centroid update step, the computation comprises
mainly separate per data point optimization problems.  The main communication
cost is on synchronizing the update for centroids by the inner loop.  
%The extra time spent on the parallelization companion is the MPI communications
%between different nodes.  
The synchronization time with equally partitioned data is negligible.

We experimented with discrete distributions over a vector space endowed with the
Euclidean distance as well as over a symbolic set. In the second case, a symbol-to-symbol distance matrix is provided. When applying D2-clustering to such data, the step of updating the support vectors can be skipped since the set of symbols is fixed.  In some datasets, the support vectors in the distributions  locate only on a pre-given grid. We can save memory in the implementation by storing the indices of the grid points rather than the direct vector values. 

Although we assume each instance is a single distribution in all the previous discussion, it is straightforward to generalize to the case when an instance is an array of distributions (indeed that is the original setup of D2-clustering in \cite{li2008real}). For instance, a protein sequence can be characterized by three histograms over respectively amino acids, dipeptides, and tripeptides.   This extension causes little extra work in the algorithms. When updating the cluster centroids, the distributions of different modalities can be processed separately, while in the update of cluster labels, the sum of squared Wasserstein distances for all the distributions is used as the combined distance. 

\section{Complexity and Performance Comparisons}\label{sec:compl}
Recall some notations: $\bar{N}$ is the data size (total number of
distributions to be clustered); $d$ is the dimension of the support vectors;
$K$ is the number of clusters; and $\epsilon_i$ or $\epsilon_o$ is the number of iterations in the inner or outer loop.  
Let $\bar{m}$ be the average number of support
vectors in each distribution in the training set and $m$ be the number of
support vectors in each centroid distribution ($\bar{m}=m$ in our setup).

%In a stringent memory allocation scheme, we only need to store the data and
%centroid distributions throughout, and dynamically allocate space for the
%optimization variables. The memory requirement for optimization is then
%negligible comparing with data. The storage of both input and output data is
%at the order $O(d\bar{N}\bar{m}+dKm)$.  
In our implementation, to reduce the time of dynamic memory allocation, 
we retain the memory for the matching weights between the support vectors of each distribution and its
corresponding centroid.  Hence, the memory allocation is of order $O(\bar{N}\bar{m}m) + O(d\bar{N}\bar{m}+dKm)$.

For computational complexity, first consider the
time for assigning cluster labels in the outer loop.  Without the
acceleration yielded from the triangle inequality, the complexity
is $O(\epsilon_o \bar{N} K l(\bar{m}m, d))$, where $l(\bar{m}m, d)$ is the average
time to solve the Wasserstein distance between distributions on a $d$
dimensional metric space.  Empirically, we found that by 
omitting unnecessary distance computation via
the triangle inequality, the
complexity is reduced roughly to $O(\epsilon_o(\bar{N} + K^2)l(\bar{m}m, d))$.  
For the centroid update step, the time complexity of the serial
version of the ADMM method is $O(\epsilon_o \epsilon_i
\bar{N} m d) + O(T_{admm} \cdot \epsilon_o \epsilon_i \bar{N}
q(\bar{m} m,d))$, where $q(m'm,d)$ is the average time to solve QPs
(Eq.~\eqref{eq:admmqp}). 
%The time complexity for the subgradient descent method 
%is $O(\epsilon_o \epsilon_i\bar{N} m d / \tau) + O(\epsilon_o \epsilon_i \bar{N} l(\bar{m} m,d)))$. 
The complexity of the serial B-ADMM
is $O(\epsilon_o \epsilon_i \bar{N} m d / \tau) + O( \epsilon_o \epsilon_i
\bar{N} \bar{m} m)$.  Note that in the serial algorithms, the complexity for updating centroids
does not depend on $K$, but only on data size $\bar{N}$. 
%The reason is that the
%complexity of updating a single centroid is linear in the cluster size, which
%on average decreases proportionally with $K$ at its increase.
For the parallel versions of the algorithms, the communication load
per iteration in the inner loop is\break $O(T_{admm} K m d)$ for ADMM
and $O(K m (1 + d/\tau))$ for the B-ADMM.

%\subsection{Usage Guidelines}
Both analytical and empirical studies (Section \ref{sec:profile}) show
that the ADMM algorithm is significantly slower than the other two when the
data size is large due to the many constrained QP sub-problems required.
Although the theoretical properties of convergence are better understood for
ADMM, our experiments show that B-ADMM performs well consistently 
in terms of both convergence and the quality of the clustering results.  
%The major drawback of the subgradient descent algorithm is the
%difficulty in tuning the step-size (also mentioned in~\cite{benamou2014iterative}).  

Although the preference for B-ADMM is experimentally validated, given the lack of strong theoretical results on its convergence, it is not clear-cut that B-ADMM 
can always replace the alternatives.  We were thus motivated to develop the 
subgradient descent (in our supplement) 
and standard ADMM algorithms to serve at least as yardsticks for comparison. 
We provide the following guidelines on the usage of the algorithms. 
\begin{itemize}
\item We recommend the modified B-ADMM  
as the default data processing pipeline for its scalability, 
stability, and fast performance. Large memory is assumed to be available under 
the default setting. 
\item It is known that ADMM type methods can approach the optimal solution quickly at
the beginning when the current solution is far from the optimum while the
convergence slows down substantially when the solution is in the proximity of
the optimum.  Because we always reset the Lagrangian multipliers in B-ADMM
at the beginning of every round of the inner loop and a fixed number of iterations 
are performed within the loop, our scheme does not pursue aggressively 
high accuracy for the resulting centroids at every round. 
%As remarked in\cite{boyd2011distributed}, such high accuracy is often unnecessary in machine learning applications.  
However, if the need arises for highly accurate centroids, 
we recommend the subgradient descent method that takes as initialization 
the centroids first obtained by B-ADMM. 
%Caution has to be taken for tuning the step-size parameter in this method.
%\item As analyzed above, Bregman ADMM requires substantially more memory than the other two methods. 
%Hence, on a low memory streaming platform, the subgradient descent method 
%or the ADMM method can be more suitable. When the support size is small (say less than 10), 
%it is also possible to speed-up the LP/QP solver greatly by doing some pre-computation, an established practice in control theory~\cite{tondel2003algorithm,bemporad2002explicit}. 
%This grants the subgradient descent and ADMM methods some extra edge.
%However, explicit MPC suffers from its bottleneck that it always consumes a lot memory and not scalable to many variables. 
\end{itemize}

\section{Experiments}\label{sec:expr}
We have conducted experiments to examine the convergence of the algorithms, stability, computational/memory efficiency and scalability of the algorithms, and quality of the clustering results on large data from several domains. 

\begin{table}[ht!]
\centering
\caption{Datasets in the experiments. $\bar{N}$: data size, $d$: dimension of the support vectors ("symb" for symbolic data), $m$: number of support 
vectors in a centroid, $K$: maximum number of clusters tested. An entry with the same value as in the previous row is indicated by "-". }
\begin{tabular}{c|cccc}
\hline
Data & $\bar{N}$ & $d$ & $m$ & $K$ \\ \hline 
synthetic & 2,560,000  & $\ge$16 & $\ge$32 & 256 \\ \hline
image color  & 5,000 & 3 & 8 & 10 \\
image texture  & - & - & - & - \\ \hline
%protein sequence 1-gram & 10,742 & symb. & 20 & 10 \\
%protein sequence 3-gram & - & - & 32 & - \\
%protein sequence 1,2,3-gram & - & - & - & - \\ 
%\hline
USPS digits &  11,000 & 2 & 80 & 360 \\ \hline
BBC news abstract & 2,225 & 300 & 16 & 15 \\
Wiki events abstract & 1,983 & 400 & 16 & 100 \\
20newsgroups GV  & 18,774 & 300 & 64 & 40 \\ 
20newsgroups WV &  - & 400 & 100 & - \\ \hline
\end{tabular}
\label{table:stat}
\end{table}

Table~\ref{table:stat} lists the basic information about the datasets used in our experiments. For the synthetic data, the support vectors are generated by sampling from a multivariate normal distribution and then adding a heavy-tailed noise from the student's t-distribution. The probabilities on the support vectors are perturbed and normalized samples from Dirichlet distribution with symmetric prior. We omit details due to lack of space. The synthetic data are only used to study the scalability of the algorithms. The image color or texture data are created from crawled general-purpose photographs. Local color or texture features around each pixel in an image are clustered ({\it i.e.} quantized) to yield color or texture distributions. 
%The protein sequence data are histograms over the amino acids (1-gram) and tripeptides (3-tuples, 3-gram).  
The USPS digit images are treated as normalized histograms over the pixel locations covered by the digits, 
where the support vector is the 2D coordinate of a 
pixel and the weight corresponds to pixel intensity. 
For the {\em 20newsgroups} data, we use the recommended ``bydate'' MATLAB version which includes 18,774 documents and 61,188 unique words.  The two datasets, ``20 newsgroup GV'' and ``20newsgroup WV'' are created by characterizing the documents in different ways. 
The ``BBC news abstract'' and ``Wiki events abstract'' datasets are 
truncated versions of two document collections~\cite{greene2006practical,wu2015storybase}. These two sets of short documents retain only the title and the first sentence of each original post.  
The purpose of using these severely cut documents is to investigate a more challenging setting for existing document or sentence analysis methods, where semantically related sentences are less likely to share the exact same words. 
For example, ``NASA revealed its ambitions that humans can set foot on Mars''
and ``US is planning to send American astronauts to Red Planet'' describe the same event. 
More details on the data are referred to Section \ref{sec:effect}. 

\subsection{Convergence Analysis}\label{sec:convergence}
We empirically test the convergence and stability of the three approaches: modified B-ADMM, ADMM, and subgradient descent method, based on their sequential versions implemented in the C programming language. Four datasets are used in the test: protein sequence 1-gram, 3-gram data, and the image color and texture data. In summary, the experiments show that the modified B-ADMM
method has achieved the best numerical stability with respect to hyper-parameters while keeping a comparable
convergence rate as the subgradient descent method in terms of CPU time.  To conserve space, detailed results on the study of stability are provided in Appendix B. Despite of its popularity in large-scale machine learning problems, by lifting $\bar{N}$ LPs to $\bar{N}$ QPs, the ADMM approach is much slower on large datasets than the other two approaches are. 
%In our experiments, ADMM takes more than $10$ times CPU time than that of
%subgradient descent method per iteration, and is even slower compared to
%the iteration time of Bregman ADMM.

\begin{figure*}[ht!]
\centering
\subfloat[image color]{\includegraphics[clip = true, viewport = 45 180 550 590, width=0.24\textwidth]{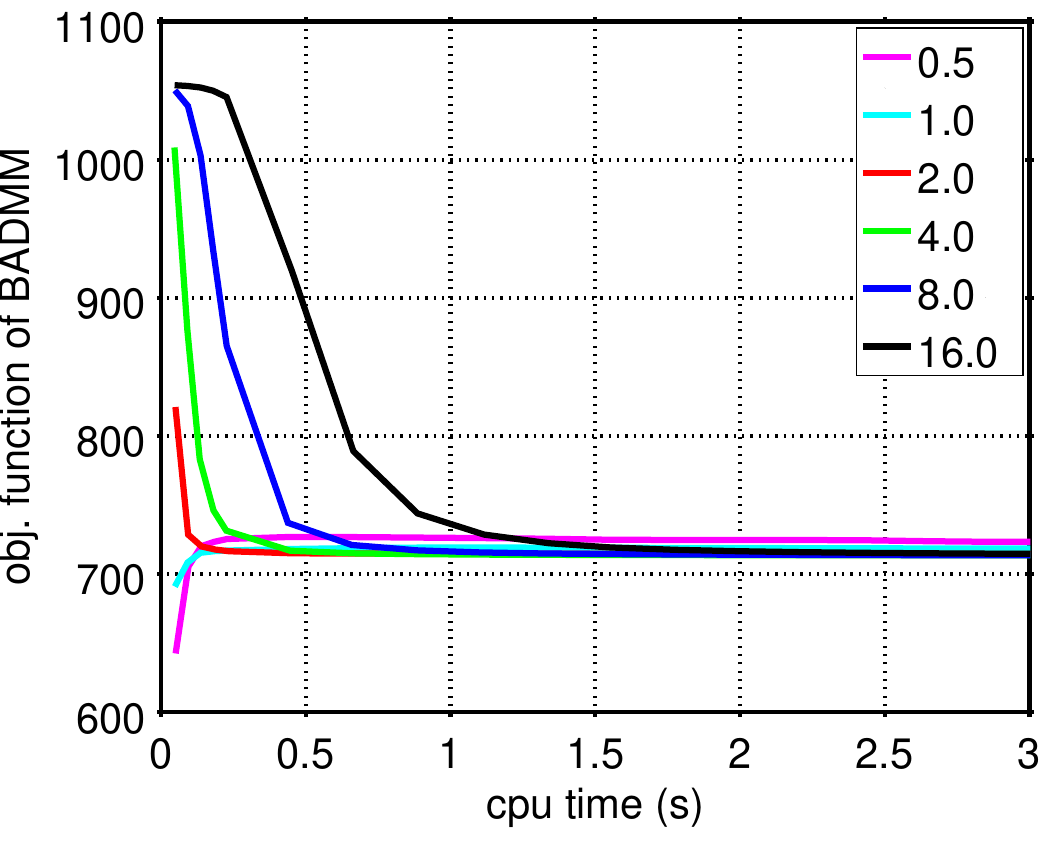}}~
\subfloat[image texture]{\includegraphics[clip = true, viewport = 45 180 550 590, width=0.24\textwidth]{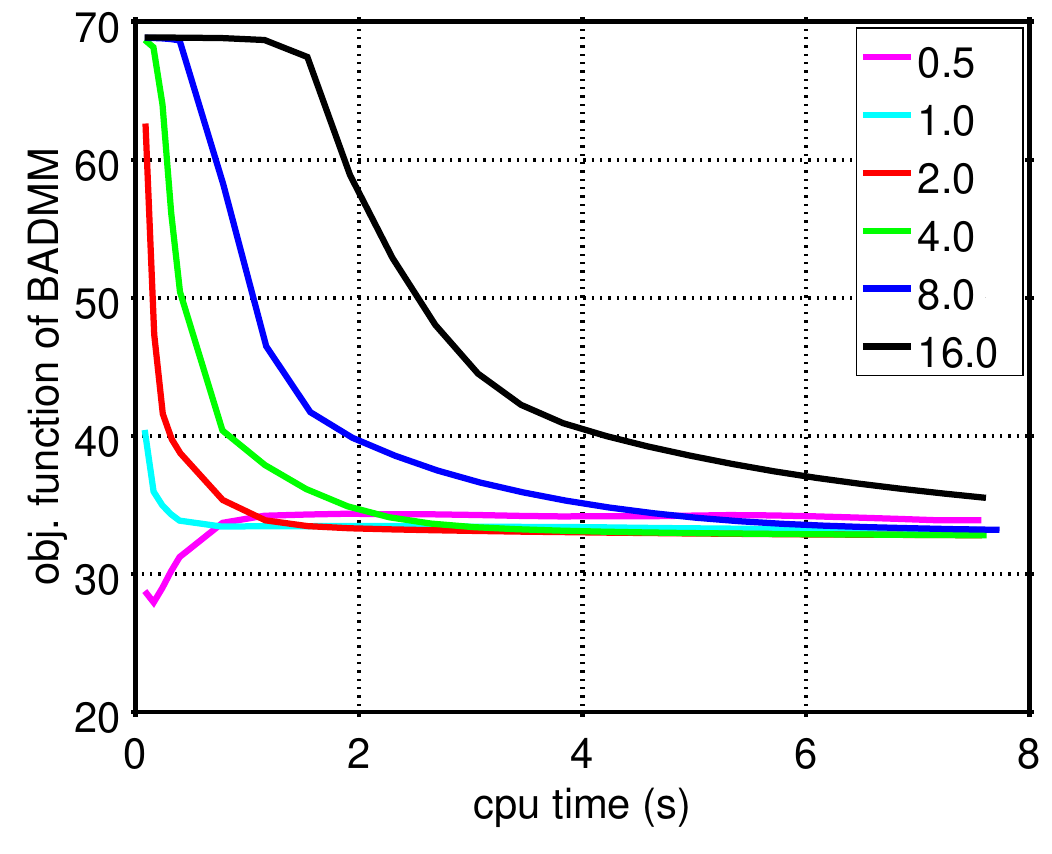}}~
\subfloat[prot. seq. 1-gram]{\includegraphics[clip = true, viewport = 45 180 550 590, width=0.24\textwidth]{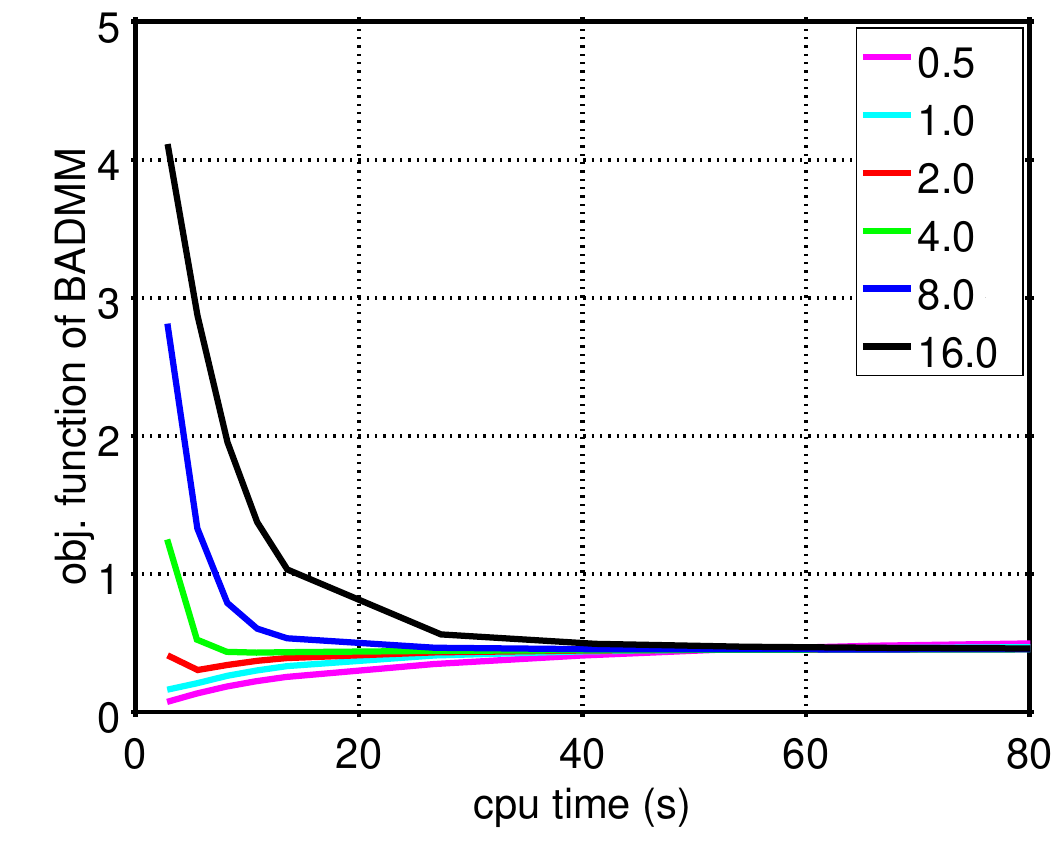}}~
\subfloat[prot. seq. 3-gram]{\includegraphics[clip = true, viewport = 45 180 550 590, width=0.24\textwidth]{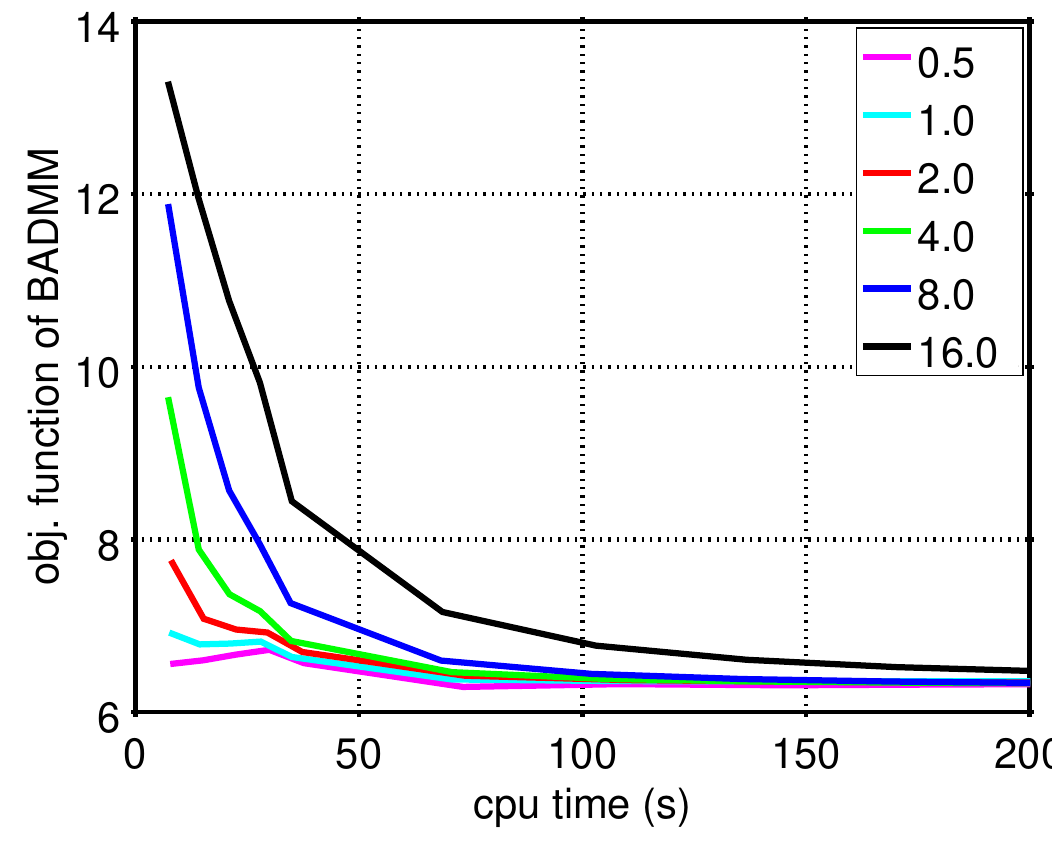}}\\
\subfloat[image color]{\includegraphics[clip = true, viewport = 45 180 550 590, width=0.24\textwidth]{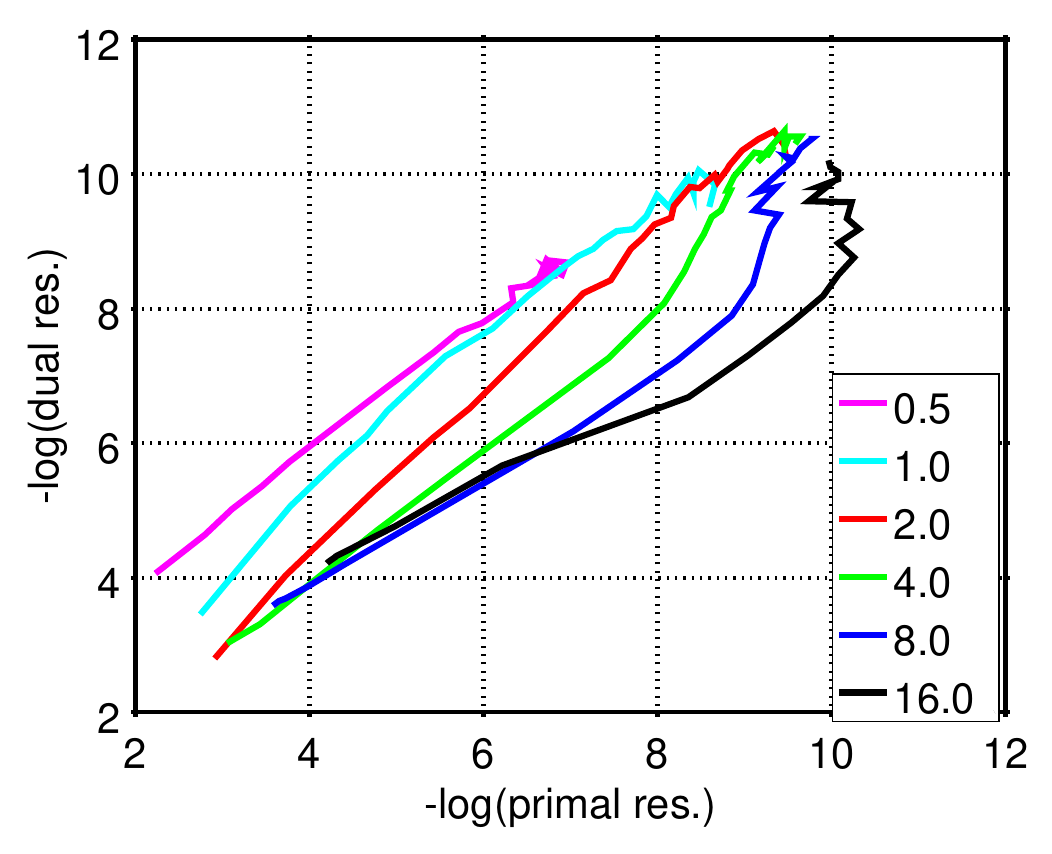}}~
\subfloat[image texture]{\includegraphics[clip = true, viewport = 45 180 550 590, width=0.24\textwidth]{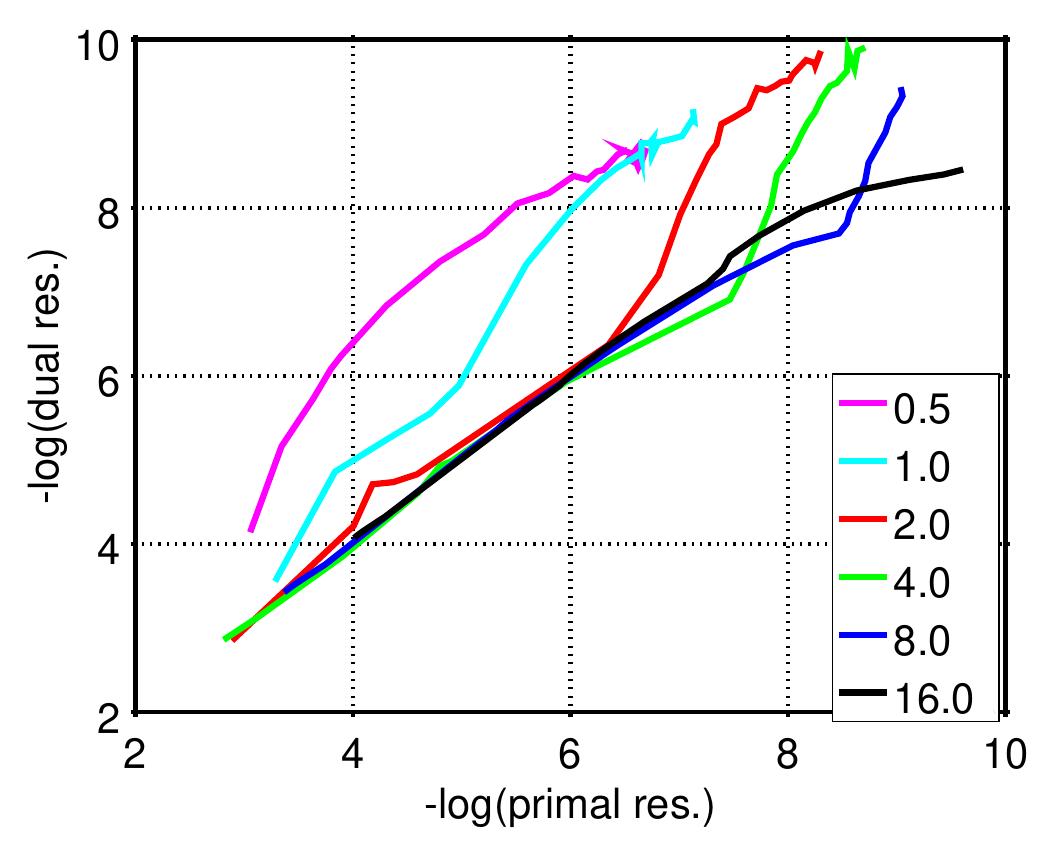}}~
\subfloat[prot. seq. 1-gram]{\includegraphics[clip = true, viewport = 45 180 550 590, width=0.24\textwidth]{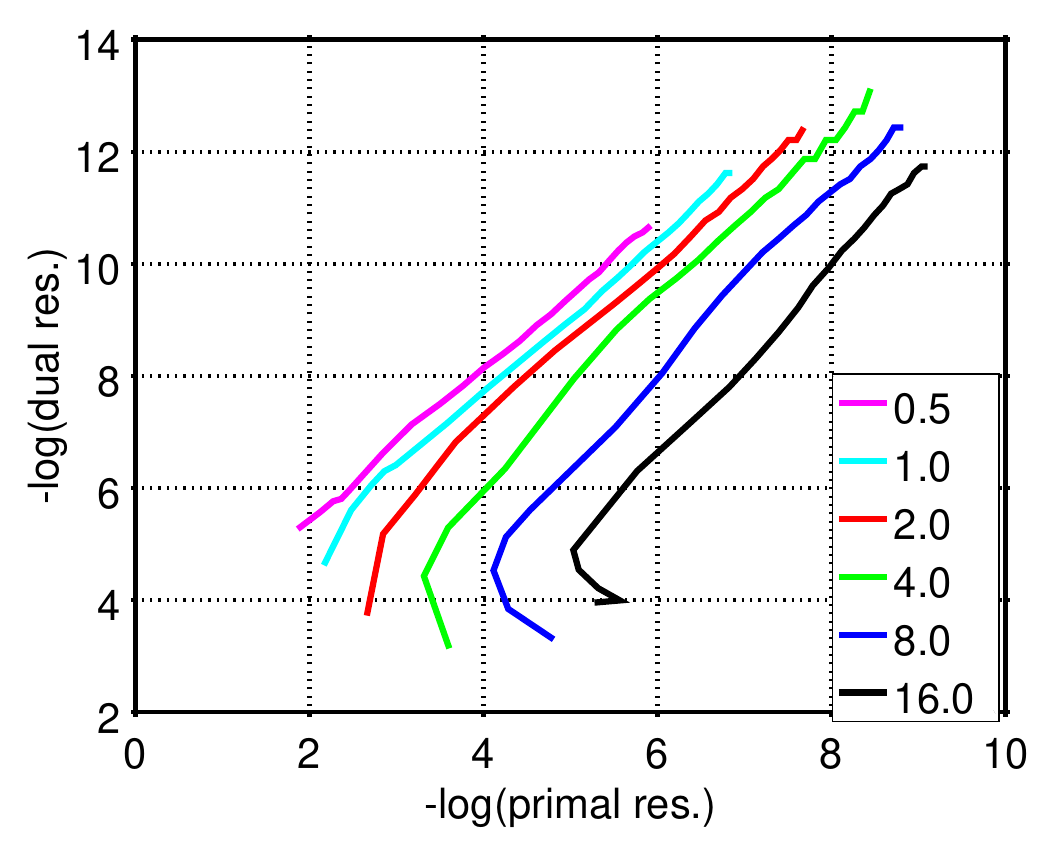}}~
\subfloat[prot. seq. 3-gram]{\includegraphics[clip = true, viewport = 45 180 550 590, width=0.24\textwidth]{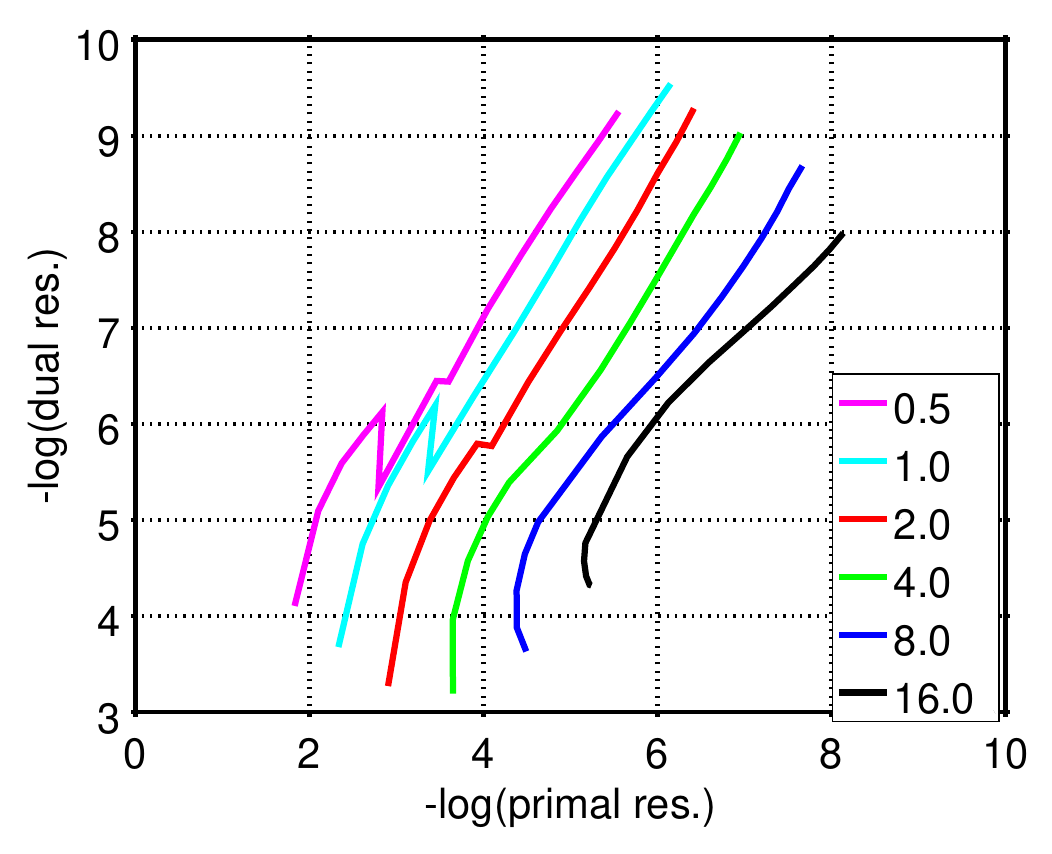}}
\caption{Convergence analysis of the B-ADMM method for computing a single centroid based on four datasets. First row: objective function of B-ADMM based centroid computation with respect to CPU time; Second row: the trajectory of 
dual residual vs. primal residual (in the negative log scale).
}
\label{fig:converge}
\end{figure*}

We examine the convergence property of the B-ADMM approach for computing the centroid of a single cluster (the inner loop). In this experiment, a subset of image color or texture data with size $2,000$ is used. For the two protein sequence datasets, the whole sets are used.
Fig.~\ref{fig:converge} shows the convergence analysis
results on the four datasets.  The vertical axis in the plots in the first row of
Fig.~\ref{fig:converge} is the objective function of B-ADMM, given in
Eq.~\eqref{eq:badmm_prob}, but not the original objective function of
clustering in Eq.~\eqref{eq:centroid}.  The runtime is based on a single thread with 2.6
GHz Intel Core i7.  The plots reveal two characteristics about the B-ADMM approach: 1)
The algorithm achieves consistent and comparable convergence rate under a wide
range of values for the hyper-parameter $\rho_0 \in \{0.5, 1.0, 2.0, 4.0,
8.0, 16.0\}$ and is numerically stable; 2) The effect of
the hyper-parameter on the decreasing ratio of the dual and primal residuals
follows similar patterns across the datasets.

\subsection{Efficiency and Scalability}\label{sec:profile}
We now study the computational/memory efficiency and scalability of AD2-clustering with the
B-ADMM algorithm embedded for computing cluster centroids. We use the synthetic data that allow easy control over data size and other parameters in order to test their effects on the computational and memory load ({\it i.e.}, workload) of the algorithm. 
%We first investigate the workload on a single thread. Once we gain a good understanding about the serial workload, 
We study the {\em scalability} of our parallel implementation on a cluster computer with distributed memory. Scalability here refers to the ability of a
parallel system to utilize an increasing number of processors.

% \textbf{Serial Experiments:} 
% %Given the complexity of the algorithm, it is not straightforward to quantify  
% %the relationship between workload and the data parameters . 
% To reveal the dependence of workload on any particular parameter (see the definitions of $\bar{N}$, $K$, $d$ and $m$ in Section~\ref{sec:compl}), we fixed the other parameters at a set of representative values, and monitored memory and time consumed in one outer-loop iteration. 
% We set $\bar{N}=5000$, $K=16$, $d=16$, $m=16$ as the baseline values.
% %with 50 seeds to generate the synthetic data. 
% We run each task subject to the 
% same parameters three times and took the median values of memory and time required. 
% We found the following empirical trends. Both time and memory scale roughly linearly with $\bar{N}$. The effect of $d$ upon time or memory is small when
% $d$ is small (say $d\le m$), but larger $d$ yields roughly linear increase in time and memory. Workload in both time and memory scale almost quadratically with $m$. Time scales linearly with $K$, while memory is affected little by $K$. 

%\textbf{Parallel Experiments:} 
AD2-clustering can be both CPU-bound and memory-bound. Based on the observations from 
the above serial experiments, we conducted three sets of experiments to 
test the scalability of AD2-clustering 
in a multi-core environment, specifically, strong scaling efficiency, 
weak scaling efficiency with respect to $\bar{N}$ or $m$.
The configuration ensures that each iteration finishes within one hour
and the memory of a typical computer cluster is sufficient. 

\begin{table}[ht!]\small
\centering
\caption{Scaling efficiency of AD2-clustering in parallel implementation.}
\begin{tabular}{c|ccccc}
\hline
\# processors &  32 & 64 & 128 & 256 & 512 \\\hline
SSE ($\%$) & 93.9 & 93.4 & 92.9 & 84.8 & 84.1 \\
WSE on $\bar{N}$ ($\%$) & 99 & 94.8 & 95.7 & 93.3 & 93.2\\
WSE on $m$ ($\%$) & 96.6 & 89.4 & 83.5 & 79.0 & - \\ \hline
\end{tabular}
\label{table:scale}
\end{table}
{\em Strong scaling efficiency (SSE)} is about the speed-up gained from using more and more processors when the problem is fixed in size. Ideally, the runtime on parallel CPUs is the time on a single thread divided by the number of CPUs.  In practice, such a reduction in time cannot be fully achieved due to communication between CPUs and time for synchronization. We thus measure SSE by the ratio between the ideal and the actual amount of time. We chose a moderate size problem 
that can fit in the memory of a single machine (50GB):
$\bar{N}=250000$, $d=16$, $m=64$, $k=16$.   Table~\ref{table:scale} shows
the SSE values with the number of processors
ranging from 32 to 512.  The results show that AD2-clustering scales well in SSE 
when the number of processors is up to hundreds. 

{\em Weak scaling efficiency (WSE)} measures how stable the real computation time 
can be when proportionally more processors are used as the size of the problem grows.
We compute WSE with respect to both $\bar{N}$ and $m$. 
Let $np$ be the number of processors. For WSE on $\bar{N}$, 
we set $\bar{N}=5000 \cdot np$, $d=64$, $m=64$, and $K=64$ on each processor. The per-node memory is 
roughly $1GB$. For WSE on $m$, we 
set $\bar{N}=10000$, $K=64$, $d=64$, and $m=32 \cdot \sqrt{np}$. 
Table~\ref{table:scale} shows the values of WSE on $\bar{N}$ and $m$. 
We can see that AD2-clustering also has good weak scalability, making it suitable for handling large scale data.  In summary, our proposed method can be effectively accelerated with an increasing number of CPUs. 

%In comparison, the previous work by Zhang et al.~\cite{zhang2015parallel} can
%hardly achieve any further speed-ups when the number of cores exceeds 32. 
%% Comment: the remark is not really true. Maybe Zhang's implementation does not
%% scale well because of some fixed parameters, but in principle should be able
%% to scale well.

\subsection{Quality of Clustering Results}\label{sec:effect}
\begin{figure}[ht!]
\subfloat[Homogeneity vs. completeness]{
    \includegraphics[ viewport = 45 180 550 590,clip,width=0.235\textwidth]{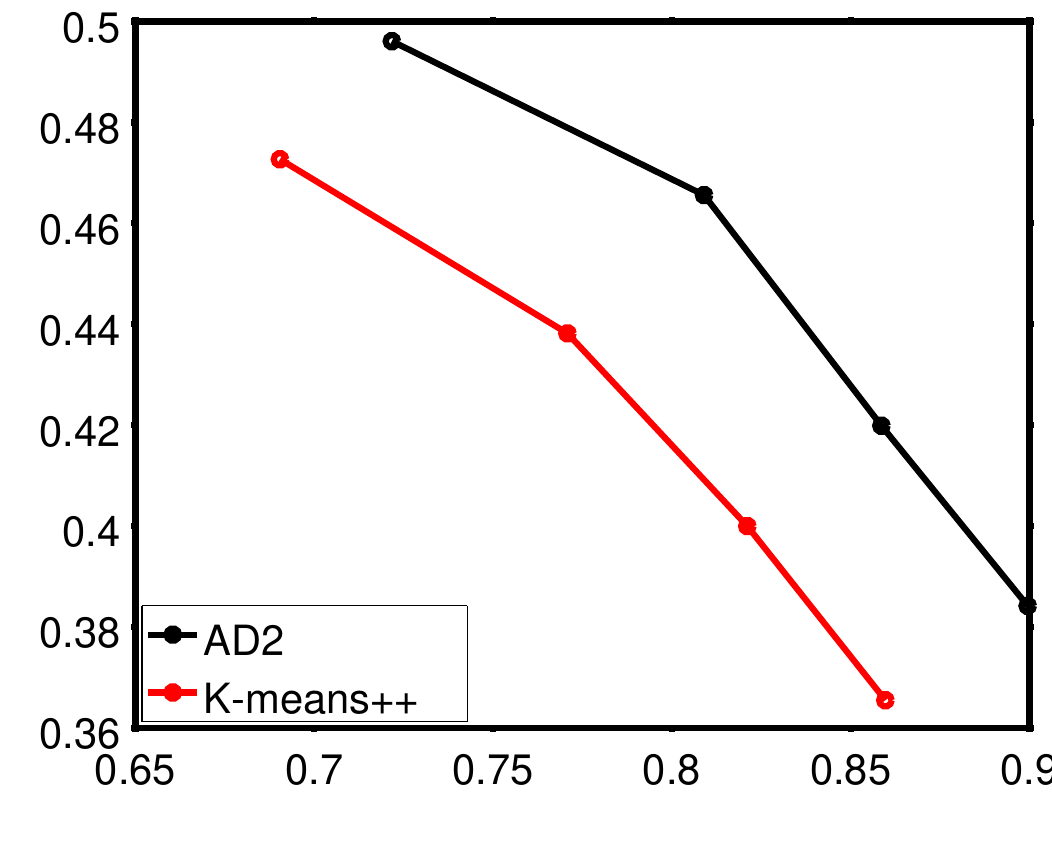}
    \label{fig:usps-clu}
  }
\subfloat[Test error rate vs. blankout rate]{
    \includegraphics[ viewport = 45 180 550 590,clip,width=0.235\textwidth]{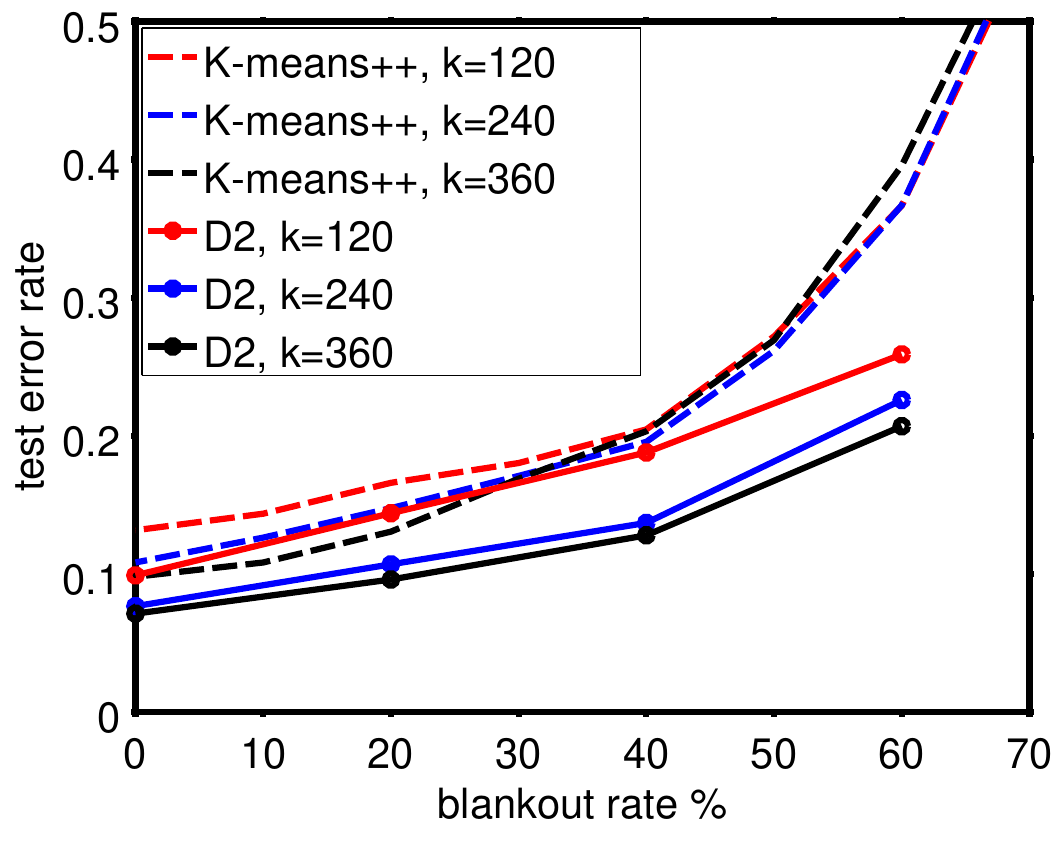}
    \label{fig:usps-cla}
  }
  \caption{Comparisons between Kmeans++ and AD2-clustering on USPS dataset. 
  We empirically set the number of support vectors in the centroids
$m=80 (1-blankout\_rate)$.}
\end{figure}

%In the clustering of the three real world datasets below, no extra parameters regarding the optimization procedure are needed to be \textit{ad-hoc} specified in our implementation.
\textbf{Handwritten Digits:}
We conducted two experiments to
evaluate the results of AD2-clustering on USPS data, which contain 
$1100\times 10$ instances ($1,100$ per class). 
First, we cluster the images at $K=30, 60, 120, 240$
and report in Figure~\ref{fig:usps-clu} the homogeneity versus completeness~\cite{rosenberg2007v} of the obtained clustering results. 
We set $K$ to large values because clustering performed on such image data is often for the purpose of quantization where the number of clusters is much larger than the number of classes. In this case, homogeneity and completeness are more meaningful measures than the others used in the literature (several of which will be used later for the next two datasets). Roughly speaking, completeness measures how likely members of the same true class fall into the same cluster, while  homogeneity measures how likely members of the same cluster belong to the same true class. By construction, the two measures have to be traded off. 
We compared our method with Kmeans++~\cite{arthur2007k}. For this dataset, we found that Kmeans++, with more careful initialization, yields better results than the standard K-means.  Their difference on the other datasets is negligible. Figure~\ref{fig:usps-clu} shows that AD2-clustering obviously outperforms Kmeans++  cross $K$'s. 
%It takes about an hour to finish AD2-clustering on the whole dataset at $K=240$.  
%Our experiments show that the result of Kmeans++ varies considerably depending on the initialization. The plot in the figure uses the median values over multiple runs.  We found that the improvement of AD2-clustering over Kmeans++ exceeds one standard deviation of the Kmeans++ results. 

Secondly, we tested AD2-clustering for quantization with the existence of
noise. In this experiment, we corrupted each sample by "blankout"---randomly deleting a percentage of pixels occupied by the digit (setting to zero 
the weights of the corresponding bins), as is done in~\cite{globerson2006nightmare}.  Then each class is randomly split into 800/300 training and test samples. Clustering
is performed on the 8000 training samples; and a class label is assigned
to each cluster by majority vote.  In the testing phase, classifying an
instance entails locating its nearest centroid and then assigning the class label of the
corresponding cluster.  The test classification
error rates with respect to $K$ and the blankout rate are plotted in Figure~\ref{fig:usps-cla}. The comparison with Kmeans++ demonstrates that AD2-clustering performs consistently better, and the margin is remarkable when the number of clusters is large and the blankout rate is high. 

\begin{table}[htp]
% \begin{minipage}[t]{0.67\textwidth}\vspace{0pt}
\caption{Compare clustering results of AD2-clustering and several baseline
  methods using two versions of Bag-of-Words
  representation for the 20newsgroups data. Top panel: the data are extracted
  using the GV vocabulary; bottom panel: WV vocabulary. AD2-clustering is performed once on 16 cores with less than 5GB memory. Run-times of AD2-clustering are reported (along with the total number of iterations).}\label{tab:doc}
\begin{tabular}{c|cccccccccc}
\hline
GV  & \multirow{2}{*}{tf-idf}  & \multirow{2}{*}{LDA} & LDA & Avg. & \multirow{2}{*}{AD2} & \multirow{2}{*}{AD2} & \multirow{2}{*}{AD2}\\
Vocab. &  &  & na\"ive & vector & & & \\\hline
$K$ & 40 & 20 & 20 & 30 & 20 & 30 & 40 \\ 
AMI & 0.447 & 0.326 & 0.329 & 0.360 & 0.418 & \textbf{0.461} & 0.446 \\
ARI & 0.151 & 0.160 & 0.187 & 0.198 & 0.260 & 0.281 & \textbf{0.284} \\
hours &&&&& 5.8 & 7.5 & 10.4 \\
\# iter.&&&&& 44 & 45 & 61\\ \hline 
\end{tabular}
\vskip 0.1in
\begin{tabular}{c|cccccccccc}
\hline
WV  & \multirow{2}{*}{tf-idf}  & \multirow{2}{*}{LDA} & LDA & Avg. & \multirow{2}{*}{AD2} & \multirow{2}{*}{AD2} & \multirow{2}{*}{AD2}\\
Vocab. &  &  & na\"ive & vector & & & \\\hline
%WV Voc. & tf-idf  & LDA & LDA+Naive & Ave. vec. & AD2 & AD2 & AD2\\\hline
$K$ & 20 & 25 & 20 & 20 & 20 & 30 & 40 \\
AMI & 0.432 & 0.336 & 0.345 & 0.398 & 0.476 & \textbf{0.477} & 0.455\\
ARI & 0.146 & 0.164 & 0.183 & 0.212 & \textbf{0.289} & 0.274 & 0.278 \\
hours &&&&& 10.0 & 11.3 & 17.1 \\
\# iter. &&&&& 28 & 29 & 36\\ \hline
\end{tabular}
%\end{minipage}\hfill
%\begin{minipage}[t]{0.3\textwidth}
%\end{minipage}
\end{table}

\textbf{Documents as Bags of Word-vectors:}
The idea of treating each document as a bag of vectors has been explored
in previous work where a nearest neighbor classifier 
is constructed using Wasserstein distance~\cite{wan2007novel,kusner2015word}. 
One advantage of the Wasserstein distance is to account for the many-to-many mapping between
two sets of words. 
However, clustering based on Wasserstein distance, especially the use of 
Wasserstein barycenter, has not been explored in the literature of document analysis. 
We have designed two kinds of experiments using different document data to assess the power of AD2-clustering.

To demonstrate the robustness of D2-clustering across different word embedding spaces, 
we use 20newsgroups processed based on two pre-trained word embedding models.
We pre-processed the dataset by two steps:
remove stop words and remove other words that do not belong to a pre-selected
background vocabulary. In particular, two background vocabularies are tested: 
English Gigaword-5 (denoted by GV) \cite{pennington2014glove} and
a Wikipedia dump with minimum word
count of 10 (denoted by WV) \cite{mikolov2013distributed}.
%We experimented with applying the SVM classifier to tf-idf word frequency vectors using the original vocabulary (only stop words removed) and the GV and WV vocabularies. The classification error rates obtained are quite close,
%all around $19\%$, differing only by about $1\%$ from each other.  This 
%indicating that 
Omitting details due to lack of space, we validated that under the GV or WV vocabulary
information relevant to the class identities of the documents is almost intact.
The words in a document are then mapped to a vector space. 
%The document analysis community has found the mapping useful for capturing between word similarity and promoted its use. 
For GV vocabulary, the Glove mapping to a vector space of dimension $300$ is used~\cite{pennington2014glove}, while for WV, the Skip-gram model is used to train a mapping space of dimension $400$~\cite{mikolov2013distributed}. The frequencies on the words are adjusted by the popular scheme of tf-idf.  
The number of different words in a document is bounded by
$m$ (its value in Table~\ref{table:stat}).  If a document has more than
$m$ different words, some words are merged into hyper-words recursively until
reaching $m$, in the same manner as the greedy merging scheme used in
centroid initialization described in Section~\ref{sec:impl}.

We evaluate the clustering performance by two
widely used metrics: AMI\cite{vinh2010information} and
ARI\cite{rand1971objective,hubert1985comparing}.  The baseline methods for
comparison include K-means on the raw tf-idf word frequencies, K-means on the LDA
topic proportional vectors\cite{hoffman2010online} (the number of LDA topics is chosen from $\{40,60,80,100\}$), K-means on the average word
vectors, and the na\"ive way of treating the 20 LDA topics as clusters.  
For each baseline method, 
we tested the number of clusters $K\in\{10,15,20,25,30,40\}$ and report only the best performance for the baseline methods in Table~\ref{tab:doc}, while for AD2-clustering, $K=20, 30, 40$ are reported. Under any given setup of a baseline method, multiple runs were conducted with different initialization and the median value of the results was taken. 
%All baseline methods except tf-idf
%have standard deviation smaller than 0.01 via multiple runs, and their median
%values are shown.  
The experimental results show that AD2-clustering achieves the best
performance on the two datasets according to both AMI and ARI. Comparing with most
baseline methods, the boost in performance by AD2-clustering is substantial.  
%K-means on tf-idf yields AMI close to the results of AD2-clustering, but
%the clusters generated by the former are quite unbalanced.  Unbalanced clusters
%tend to yield higher values in AMI due to the nature of AMI. 
%By ARI, however, K-means on tf-idf has no advantage over the other baseline methods. 
Furthermore, we also vary $m$ in the experimental setup of AD2-clustering. At $m=1$,
our method is exactly equivalent to K-means of the distribution means. We increased $m$ empirically
to see how the results improve with a larger $m$. We did not observe any further
performance improvement for $m\ge 64$. 

We note that the high competitiveness of AD2-clustering 
can be credited to (1) a reasonable word embedding model and (2) the bag-of-words model.
When the occurrence of words is sparse across documents, the semantic relatedness 
between different words and their compositions in a document 
plays a critical role in measuring the document
similarity. 

In our next experiment, we study AD2-clustering for short documents,
a challenging setting for almost all existing methods based on the bag-of-words representation.
The results show that the performance boost of AD2-clustering is also substantial. 
We use two datasets, one is called ``BBC news abstract'' and the other
``Wiki events abstract''. Each document is represented by only the title
and the first sentence from a news article or an event description.
Their word embedding models are same as the one used by the ``WV'' version in our previous experiment. 
The ``BBC news'' dataset contains five news categories, and ``Wiki events'' dataset contains 54 events. In the supplement materials, the raw data we used are provided. 
Clustering such short documents is more challenging due to the sparse nature of word occurrences. 
As shown by Table~\ref{table:short}, in terms of generating clusters coherent with the labeled categories or events, methods which leverage either the bag-of-words model or the word embedding model (but not both) are outperformed by AD2-clustering which exploits both. 
In addition, AD2-clustering is fast for those sparse support discrete distribution data. It takes only several minutes to finish the clustering in an 8-core machine. 

To quantify the gain from employing an effective word embedding model, we also applied AD2-clustering to a random word embedding model, where a vector sampled
from a multivariate Gaussian with 300 dimensions is used to represent a word in vocabulary. We found that the results are much worse 
than those reported in Table~\ref{table:short} for AD2-clustering. 
The best AMI for ``BBC news abstract'' is 0.187
and the best AMI for ``Wiki events abstract'' is 0.369, comparing respectively
with 0.759 and 0.545 obtained from a carefully trained word embedding model. 

\begin{table}
\caption{Best AMIs achieved by different methods on the two short document datasets.
NMF denotes for the non-negative matrix factorization method. }~\label{table:short}
\begin{tabular}{c|ccccc}\hline
& Tf-idf & LDA & NMF & Avg. vector & AD2 \\
BBC news abstract & 0.376 & 0.151 & 0.537 & 0.753 & 0.759 \\
Wiki events abstract & 0.448 & 0.280 & 0.395 & 0.312 & 0.545\\\hline
\end{tabular}
\end{table}

\section{Conclusions and Future Work}\label{sec:concl}
Two first-order methods for clustering discrete distributions under the Wasserstein distance have been developed and empirically compared in terms of speed, stability, and convergence.
The experiments identified the modified B-ADMM method as most preferable for D2-clustering in an overall sense under common scenarios. The resulting clustering tool is easy to use, requiring no tuning of optimization parameters. We applied the tool to several real-world datasets and evaluated the quality of the clustering results by comparing with the ground truth class labels. Experiments show that the accelerated D2-clustering often clearly outperforms other widely used methods in the respective domain of the data. 

One limitation of our current work is that the theoretical convergence property of the modified B-ADMM algorithm is not well understood. In addition, to expedite D2-clustering, we have focused on scalability with respect to data size rather than the number of support points per distribution. Thus there is room for further improvement on the efficiency of D2-clustering. Given that the algorithms explored in this paper have different strengths, it is of interest to investigate in the future whether they can be integrated to yield even stronger algorithms. 

\section*{Acknowledgment}
This material is based upon work supported by the National Science Foundation (NSF) under Grant Nos. CCF-0936948, ACI-1027854, and DMS-1521092. The primary computational infrastructures used were supported by the NSF under Grant Nos. ACI-0821527 (CyberStar) and ACI-1053575 (XSEDE). We also thank the reviewers and the associate editor for constructive comments and suggestions. 

\bibliography{fpd2}
\bibliographystyle{IEEEtran}

\appendix

\subsection{Subgradient Descent Method}\label{sec:descent}
We describe a subgradient descent approach in this section for the purpose of experimental comparison. 

Eq. {\eqref{eq:fullbatch-lp}}
can be casted as multi-level optimization by treating $\tmmathbf{w}$ as
policies/parameters and $\Pi$ as variables. Express $W^2(P, P^{(k)})$, the
squared distance between $P$ and $P^{( k )}$, as a function of $\tmmathbf{w}$
denoted by $\tilde{W} ( \tmmathbf{w} )^{(k)} \nosymbol$.
$\tilde{W}(\tmmathbf{w})^{(k)}$ is the solution to a designed optimization,
but has no closed form.  Let
$\tilde{W}(\tmmathbf{w})=\frac{1}{N}\sum_{k=1}^{N}\tilde{W}(\tmmathbf{w})^{(k)}$,
where $N$ is the number of instances in the cluster. Note that
Eq. {\eqref{eq:fullbatch-lp}} minimizes $\tilde{W}(\tmmathbf{w})$ up to a
constant multiplier.  The minimization of $\tilde{W}$ with respect to
$\tmmathbf{w}$ is thus a bi-level optimization problem.  In the
special case when the designed problem is LP and the parameters only appear
on the right hand side (RHS) of the constraints or are linear in the
objective, the subgradient, specifically $\nabla \tilde{W} ( \tmmathbf{w} )^{( k
  )}$ in our problem, can be solved via the same (dual) LP.

Again, we consider the routine that alternates the updates of $\{ x_{i} \}$ and
$\{ \pi_{i,j} \}^{( k )}$ iteratively. With fixed $\{ x_{i} \}$, updating $\{
\pi_{i,j} \}^{( k )}$ involves solving $N$ 
LP {\eqref{eq:primal}}. \ With LP {\eqref{eq:primal}} solved, we can write
$\nabla \tilde{W} ( \tmmathbf{w} )^{( k )}$ in closed form, which is given by the set
of dual variables $\{ \lambda^{( k )}_{i} \}_{i=1}^{m}$ corresponding to
$\left\{ \sum_{j=1}^{m_k} \pi_{i,j} =w_{i}, i=1, ..., m \right\}$. Because $\{ w_{i}
\}$ must reside in the facets defined by $\Delta_m$, the projected subgradient $\nabla \tilde{W}
( \tmmathbf{w} )^{( k )}$ is given by
\begin{equation}
  \nabla \tilde{W} ( \tmmathbf{w} )^{( k )} = ( \lambda^{( k )}_{1} , \ldots ,
  \lambda_{m}^{( k )} ) - \left( \sum_{i=1}^{m} \lambda^{( k )}_{i} \right) (
  1, \ldots ,1 ) \;. \label{eq:dualvar}
\end{equation}

In the standard method of gradient descent, a line search is conducted in each iteration to determine the step-size for a strictly descending update.  Line search however is computationally intensive for our problem because Eq. {\eqref{eq:dualvar}} requires solving a LP and we need Eq. {\eqref{eq:dualvar}} sweeping over $k=1, ..., N$.  In machine learning algorithms, one practice to avoid expensive line search is by using a pre-determined step-size, which is allowed to vary across iterations.  We adopt this approach here. 

One issue resulting from a pre-determined step-size is that the updated weight vector $\tmmathbf w$ 
may have negative components. We overcome this numerical instability by the technique of re-parametrization.
Let
\begin{equation}
w_i(\tmmathbf s) \assign \dfrac{\exp (s_i) }{\sum \exp (s_i)}\;,\; i=1, ..., m \, .
\end{equation}
We then compute the partial subgradient with respect to $s_i$ instead of $w_i$,
and update $w_i$ by updating $s_i$. Furthermore $\exp (s_i)$ are 
re-scaled in each iteration such that $\sum_{i=1}^m s_i = 0$. 

The step-size $\alpha (\tmmathbf{w})$ is chosen by
\begin{equation}
\sigma (\tmmathbf{w}) \assign  
\min \left(\dfrac{\alpha}{\|\nabla_{\tmmathbf s} \tilde{W} ( \tmmathbf{w}(\tmmathbf s) )\|}, \zeta\right)\;.
\label{eq:stepsize}
\end{equation}
The two hyper-parameters $\alpha$ and $\zeta$ trade off the convergence 
speed and the guaranteed decrease of the objective. Another hyper-parameter is $\tau$ which indicates the
ratio between the update frequency of  weights $\{w_i\}$ and that of support points $\{x_i\}$. 
In our experiments, we alternate one round of update for both $\{w_i\}$ and $\{x_i\}$.  
We summarize the subgradient descent approach in Algorithm~\ref{alg:gd}.

\begin{algorithm}[ht!]
\caption{Centroid Update with Subgradient Descent}\label{alg:gd}
\begin{algorithmic}[1]
\Procedure{Centroid}{$\{ P^{( k )} \}_{k=1}^{N}$, $P$}\Comment{with initial guess}
  \Repeat
  \State Updates $\{ x_{i} \}$ from Eq.{\eqref{eq:updatex}} Every $\tau$ iterations; 
  \For{$k=1, \ldots ,N$}
  \State Obtain $\Pi^{( k )}$ and $\Lambda^{(k )}$ from LP: $W ( P,P^{( k )} )$
  \EndFor
  \State $\nabla \tilde{W} ( \tmmathbf{w} ) \assign \frac{1}{N}
  \sum_{k=1}^{N} \nabla \tilde{W} ( \tmmathbf{w} )^{( k )}$; \Comment{See Eq.{\eqref{eq:dualvar}}}
  
  \State $s_i \assign  s_i - \sigma (\tmmathbf{w}) 
  \nabla \tilde{W} ( \tmmathbf{w} ) \cdot \dfrac{\partial \tmmathbf{w}}{\partial s_i}$
  \Comment{See Eq. \eqref{eq:stepsize}}
  
  \State $s_{i} \assign s_i - \sum_{j=1}^m s_j ,i=1, \ldots m$; \Comment{rescaling step}
  
  \State ${w_i} \assign \dfrac{\exp (s_i) }{\sum \exp (s_i)} ,i=1, \ldots m$; \Comment{sum-to-one}
  
  \Until{$P$ converges}
  \State \Return $P$
\EndProcedure
\end{algorithmic}
\end{algorithm}

If the support points $\{x_i\}$ are fixed, the centroid optimization 
is a linear programming in terms of $\{w_i\}$, thus convex. The subgradient 
descent method converges under mild conditions on the smoothness of the solution and small or adaptive step-sizes. 
In Algorithm~\ref{alg:gd}, the support points are also updated, and the problem becomes non-convex.

It is technically subtle to compare the convergence and stability of the overall AD2-clustering embedded with different algorithms for computing the centroid.  Because of the many iterations in the outer loop, the centroid computation algorithm (solving the inner loop) may behave quite differently over the outer-loop rounds.  For instance, if an algorithm is highly sensitive to a hyper-parameter in optimization, the hyper-parameter chosen based on earlier rounds may yield slow convergence later or even cause the failure of convergence. Moreover, achieving high accuracy for centroids in earlier rounds, usually demanding more inner-loop iterations, 
may not necessarily result in faster decrease in the clustering objective function
because the cluster assignment step also matters.  

In light of these issues,
we employ a protocol described in Algorithm~\ref{alg:protocol} to decide the number of iterations in the inner loop.  The protocol specifies that in each iteration of the outer loop, the inner loop for updating centroids should complete within $\eta T_a/ K$ amount of time, where $T_a$ is the time used by the assignment step and $K$ is the number of clusters.  
As we have pointed out, the LP/QP solver in the subgradient descent method or standard ADMM suffers from rapidly increasing complexity when the number of support points per distribution increases. In contrast, the effect on B-ADMM is much lower. 
In the experiment below, the datasets  contain distributions 
with relatively small support sizes (a setup favoring the former two methods).  
A relatively tight time-allocation $\eta = 2.0$ is set. The
subgradient descent method finishes at most $2$ iterations in the inner loop,
while B-ADMM on average finishes more than 60 iterations on the color
and texture data, and more than 20 iterations on the protein sequence 1-gram and 3-gram data. 
The results by the ADMM method are omitted because this method cannot finish a single iteration under this time allocation.  
\begin{algorithm}[htp]
\caption{Time allocation based algorithmic profile protocol}
\label{alg:protocol}
\begin{algorithmic}[1]
\Procedure{Profile}{$\{ P^{( k )} \}_{k=1}^{M}$, $Q^{}$, $K$, $\eta$}.
  \State Start profiling;
  \State $T = 0$;
  \Repeat
  \State $T_a = 0$;
  \State Assignment Step;    
  \State GetElaspedCPUTime($T_a$, $T$);
  \State GetAndDisplayPerplexity($T$);
  \State Update Step within CPU time allocation $\eta T_a / K$;
  \Until{$T < T_{total}$}  
  \State \Return{}
\EndProcedure
\end{algorithmic}
\end{algorithm}

\begin{figure}[ht!]
\centering
\subfloat[image color]{\includegraphics[clip = true, viewport = 45 180 550 590, width=0.24\textwidth]{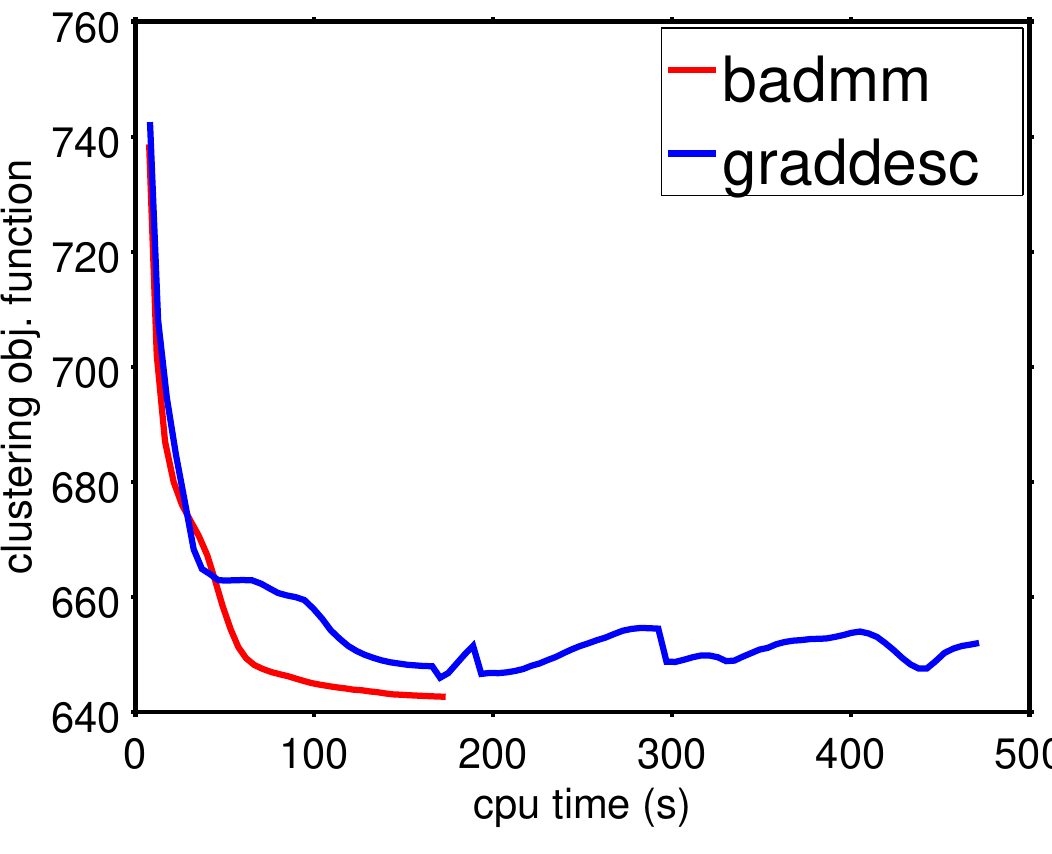}}~
\subfloat[image texture]{\includegraphics[clip = true, viewport = 50 180 550 590, width=0.24\textwidth]{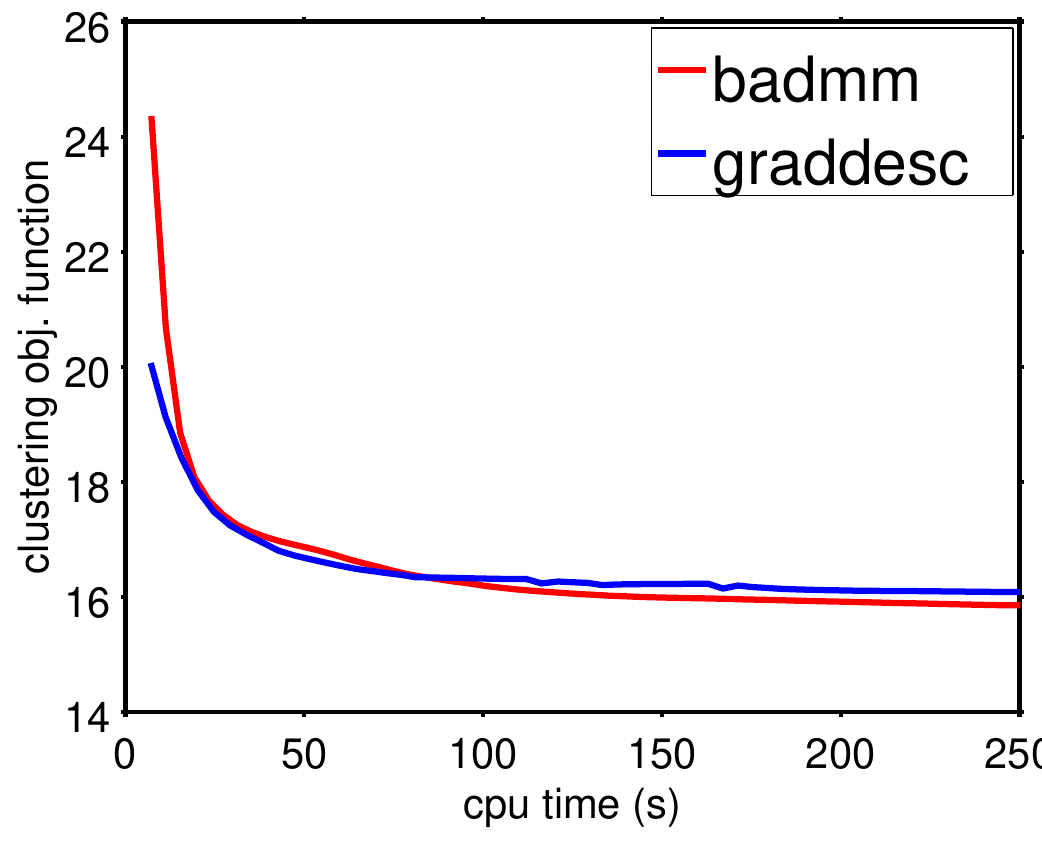}}
\\
\subfloat[prot. seq. 1-gram]{\includegraphics[clip = true, viewport = 45 180 550 590, width=0.24\textwidth]{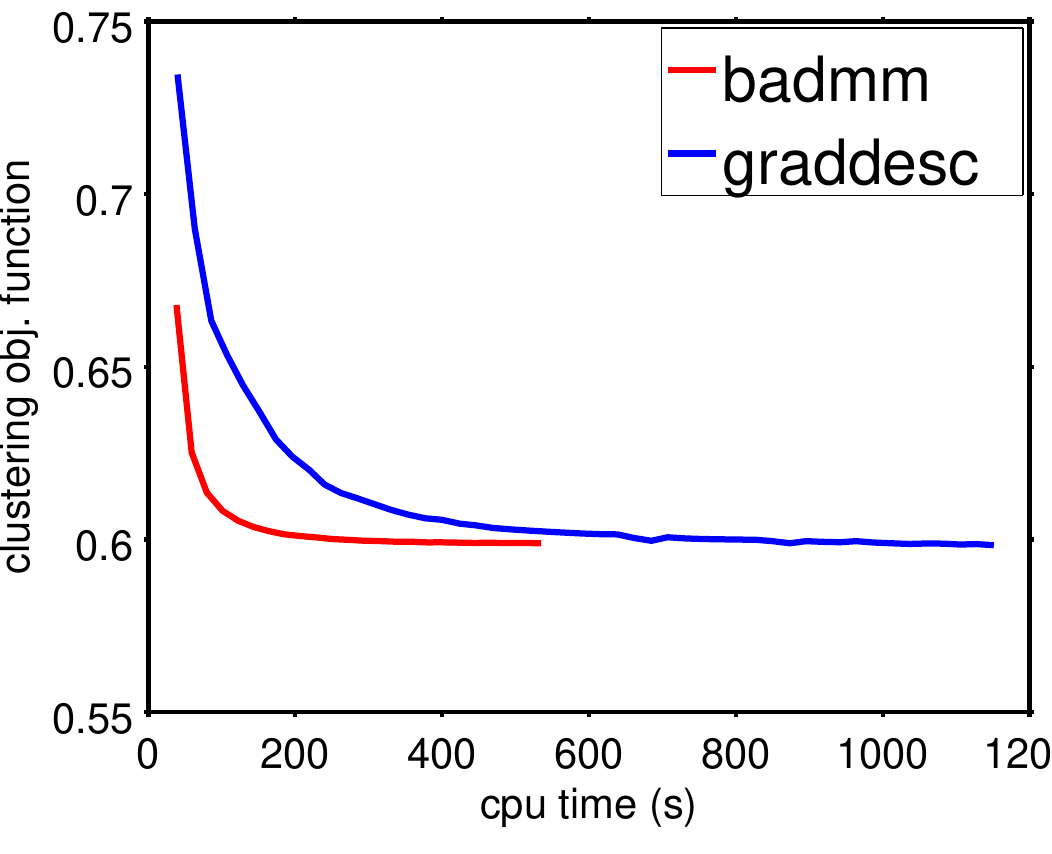}}~
\subfloat[prot. seq. 3-gram]{\includegraphics[clip = true, viewport = 45 180 550 590, width=0.24\textwidth]{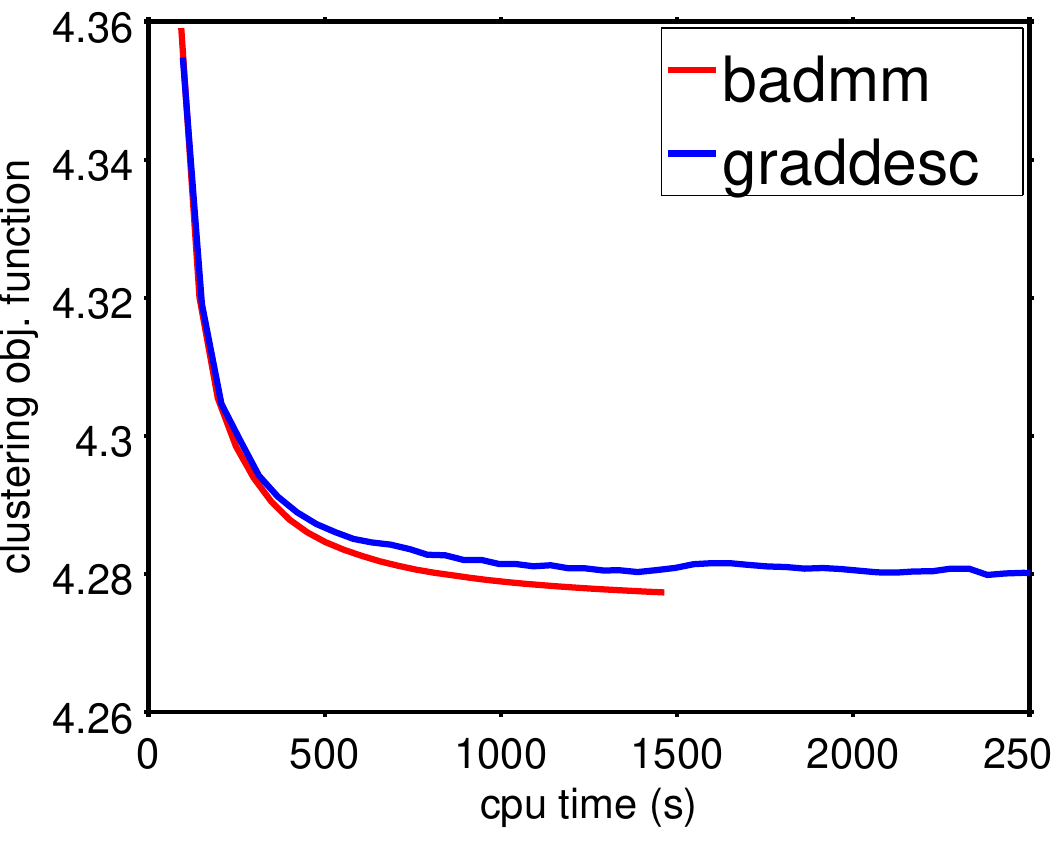}}
\caption{Convergence performance of B-ADMM and the subgradient descent
  method for D2-clustering based on four datasets. The clustering objective function versus CPU time is shown. Here, $K=10$, and the time-allocation ratio $\eta = 2.0$.}
\label{fig:allconverge}
\end{figure}
\subsection{Stability analysis of AD2-clustering}\label{sec:stability}

In Fig.~\ref{fig:allconverge}, we compare the convergence performance of the overall 
clustering process employing B-ADMM at $\rho_0=2.0$ and 
the subgradient descent method with fine-tuned values for
the step-size parameter $\alpha\in\{0.05, 0.1, 0.25, 0.5, 1.0, 2.0, 5.0, 10.0\}$. 
The step-size is chosen as the value yielding the lowest clustering objective function in the first round.  In this experiment, the whole image color and texture data are used. In the
plots, the clustering objective function (Eq.~\eqref{eq:centroid}) is shown with
respect to the CPU time. We observe a couple of advantages of the B-ADMM
method. First, with a fixed parameter $\rho_0$, B-ADMM yields good
convergence on all the four datasets, while the subgradient descent method
requires manually tuning the step-size $\alpha$ in order to achieve
comparable convergence speed. Second, B-ADMM achieves consistently
lower values for the objective function across time.  On the protein sequence 1-gram
data, B-ADMM converges substantially faster than the
subgradient descent method with a fine-tuned step-size.  Moreover, the
subgradient descent method is numerically less stable.  Although the
step-size is fine-tuned based on the performance at the beginning, on the
image color data, the objective function fluctuates noticeably in later rounds. Striking a balance (assuming it exists) between stability and speed for the subgradient descent method is a difficult dilemma.

\begin{IEEEbiography}[{\includegraphics[height=1.3in, trim={30 40 20 5}, clip,keepaspectratio]{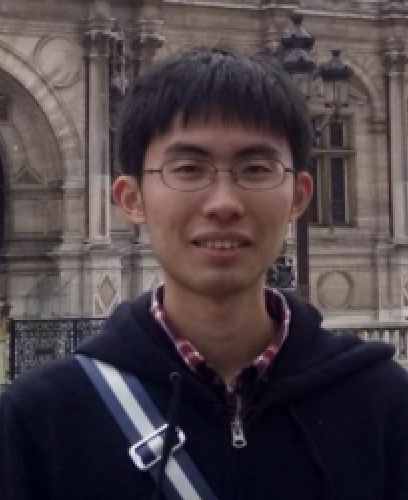}}]{Jianbo Ye} 
received his BS degree in Mathematics from the
University of Science and Technology of China in 2011. 
He worked as a research postgraduate at
The University of Hong Kong, from 2011 to 2012,
and as a research intern at Intel Labs, China in 2013. 
He is currently a PhD candidate and Research Assistant at the 
College of Information Sciences and Technology,
The Pennsylvania State University.
His research interests include statistical modeling
and learning, numerical optimization and method, and human computation for interdisciplinary studies.
\end{IEEEbiography}

\begin{IEEEbiography}[{\includegraphics[height=1.3in, trim={30 40 30 5}, clip,keepaspectratio]{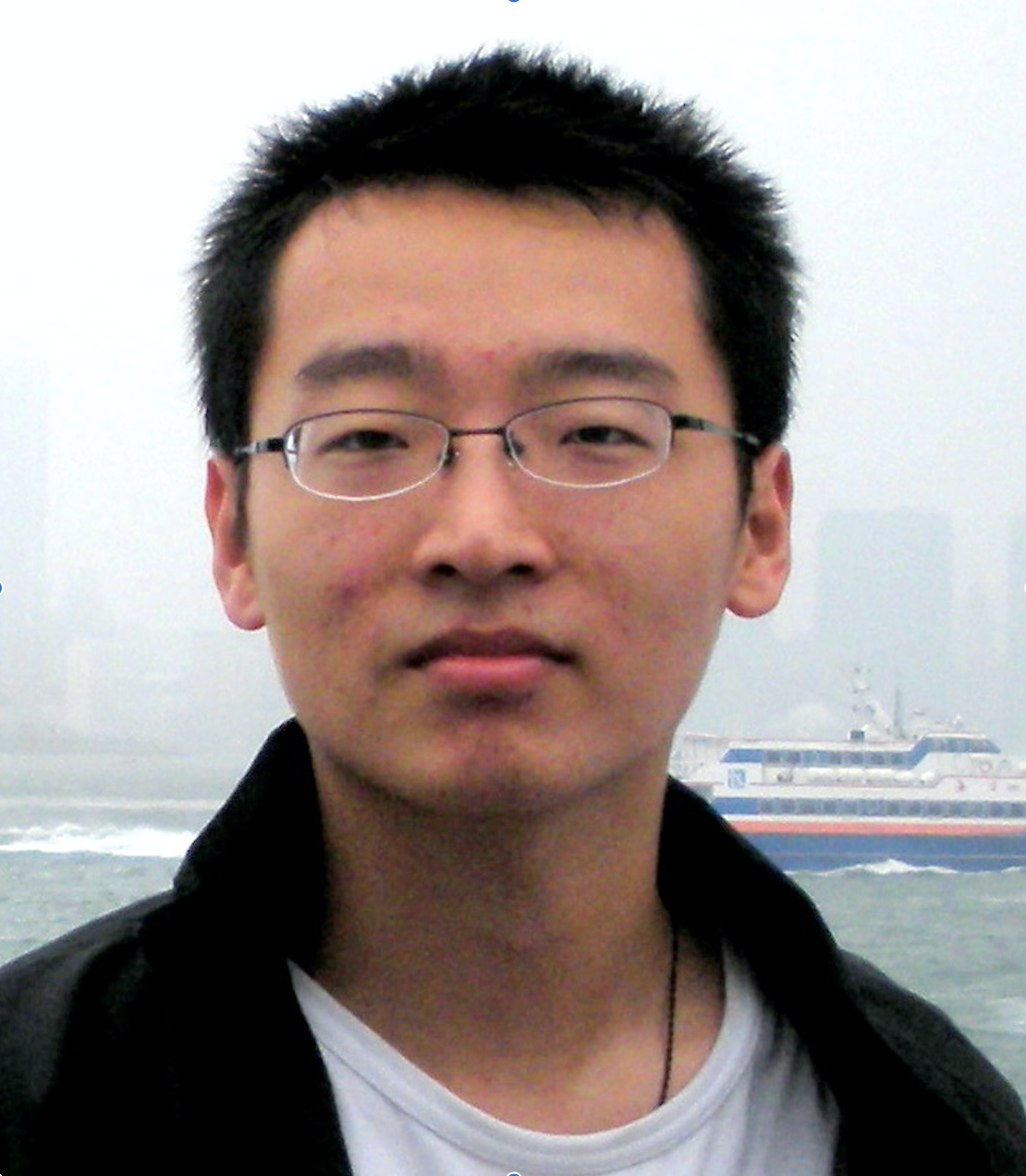}}]{Panruo Wu} 
received his BS degree in Mathematics from the University of Science and Technology of China in 2011. He is now a PhD candidate in the Supercomputing Laboratory (SuperLab) at the University of California, Riverside. He is interested in high performance computing, algorithm-based fault tolerance, and power and energy efficient computing. Wu received a student travel grant from IPDPS'14 and he was a student volunteer for ICS'15 and SC'16. He did an internship at Oak Ridge National Laboratory in 2013, an internship at Amazon in 2014, and an internship at Los Alamos National Laboratory in 2015-2016.
\end{IEEEbiography}

\begin{IEEEbiography}[{\includegraphics[width=1.2in,height=1.3in, clip,keepaspectratio,trim={0 0 0 0}]{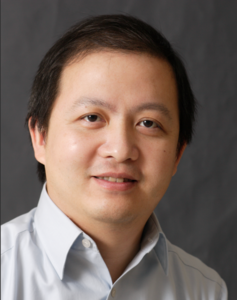}}]{James Z. Wang} is a Professor of Information Sciences and Technology at The Pennsylvania State University. He received the bachelor's degree in mathematics and computer science {\it summa cum laude} from the University of Minnesota, and the MS degree in mathematics, the MS degree in computer science and the PhD degree in medical information sciences, all from Stanford University. His research interests include computational aesthetics and emotions, automatic image tagging, image retrieval, and computerized analysis of paintings. He was a visiting professor at the Robotics Institute at Carnegie Mellon University (2007-2008), a lead special section guest editor of the IEEE Transactions on Pattern Analysis and Machine Intelligence (2008), and a program manager at the Office of the Director of the National Science Foundation (2011-2012). He was a recipient of a National Science Foundation Career award (2004).
\end{IEEEbiography}

\begin{IEEEbiography}[{\includegraphics[width=1.2in,height=1.3in, trim={0 35 20 0}, clip,keepaspectratio]{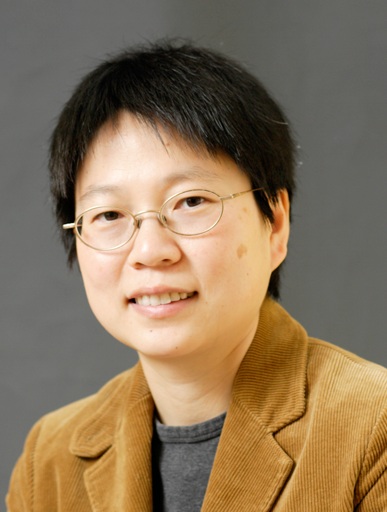}}]{Jia Li} is a Professor of Statistics at The Pennsylvania State University. She received the MS degree in Electrical Engineering, the MS degree in Statistics, and the PhD degree in Electrical Engineering, all from Stanford University. She worked as a Program Director in the Division of Mathematical Sciences at the National Science Foundation from 2011 to 2013, a Visiting Scientist at Google Labs in Pittsburgh from 2007 to 2008, a researcher at the Xerox Palo Alto Research Center from 1999 to 2000, and a Research Associate in the Computer Science Department at
Stanford University in 1999. Her research interests include statistical modeling and learning, data mining, computational biology, image processing, and image annotation and retrieval.
\end{IEEEbiography}

\vfill


\end{document}